\documentclass[useAMS,usenatbib]{mn2e}
\usepackage{longtable,lscape}
\usepackage{amsmath}
\usepackage{multirow}

\usepackage{graphicx}
\usepackage{aas_macros}
\usepackage{amssymb}

\newcommand{\vpk}{$v_\textrm{pk}$}
\newcommand{\fpk}{$F_\textrm{pk}$}
\newcommand{\lpk}{$L_\textrm{pk}$}
\newcommand{\ftot}{$F_\textrm{tot}$}
\newcommand{\ltot}{$L_\textrm{tot}$}
\newcommand{\hal}{H$\alpha$}
\newcommand{\mpk}{$M_\textrm{pk}(V)$}
\newcommand{\mplat}{$M_\textrm{plat}(V)$}

\newcommand{\about}{$\sim\!\!$~}
\newcommand{\kms}{km~s$^{-1}$}

\newcommand{\msun}{M$_\odot$}

\newcommand{\bvri}{\protect\hbox{$BV\!RI$} }

\mathchardef\mhyphen="2D

\newcommand{\be}{\begin{displaymath}}
\newcommand{\ee}{\end{displaymath}}

\def\lsim{\hbox{\rlap{\raise 0.425ex\hbox{$<$}}\lower 0.65ex\hbox{$\sim$}}}
\def\gsim{\hbox{\rlap{\raise 0.425ex\hbox{$>$}}\lower 0.65ex\hbox{$\sim$}}}

\newcommand{\ion}[2]{#1$\;${\small{#2}}\relax}

\graphicspath{{/Users/jsilverman/iip_late/plots/}}

\title[Late-Time SNe~IIP Spectra]{After the Fall: Late-Time
  Spectroscopy of Type IIP Supernovae} 

\author[Silverman, et~al.]{Jeffrey~M.~Silverman,$^{1,2}$ Stephanie Pickett,$^{1}$ J. Craig Wheeler,$^{1}$ 
\newauthor 
Alexei~V.~Filippenko,$^{3}$ J\'{o}zsef Vink\'{o},$^{1,4,5}$ G. H. Marion,$^{1,6}$ S. Bradley Cenko,$^{7,8}$ 
\newauthor 
Ryan Chornock,$^{9}$ Kelsey I. Clubb,$^{3}$ Ryan J. Foley,$^{10}$ Melissa L. Graham,$^{3,11}$ 
\newauthor 
Patrick L. Kelly,$^{3}$ Thomas Matheson,$^{12}$ Joseph C. Shields$^{9}$ \\
$^{1}$Department of Astronomy, University of Texas at Austin, Austin, TX 78712, USA \\
$^{2}$email: jsilverman@astro.as.utexas.edu \\
$^{3}$Department of Astronomy, University of California, Berkeley, CA 94720-3411, USA \\
$^{4}$Department of Optics and Quantum Electronics, University of Szeged, D\'{o}m t\'{e}r 9, 6720 Szeged, Hungary \\
$^{5}$Konkoly Observatory, Research Centre for Astronomy and Earth Sciences, Hungarian Academy of Sciences, Budapest, Hungary \\
$^{6}$Harvard-Smithsonian Center for Astrophysics, Cambridge, MA 02138, USA \\
$^{7}$Astrophysics Science Division, NASA Goddard Space Flight Center, Greenbelt, MD 20771, USA \\
$^{8}$Joint Space-Science Institute, University of Maryland, College Park, MD 20742, USA \\
$^{9}$Astrophysical Institute, Department of Physics and Astronomy, Ohio University, Athens, OH 45701, USA \\
$^{10}$Department of Astronomy and Astrophysics, University of California, Santa Cruz, CA 95064, USA \\
$^{11}$Department of Astronomy, University of Washington, Seattle WA 98195-1580, USA \\
$^{12}$National Optical Astronomy Observatory, Tucson, AZ 85719-4933, USA \\
}

\begin{document}
\date{Accepted  . Received   ; in original form  }
\pagerange{\pageref{firstpage}--\pageref{lastpage}} \pubyear{2016}
\maketitle
\label{firstpage}

\begin{abstract}
Herein we analyse late-time (post-plateau; $103 < t < 1229$~d)
optical spectra of 
low-redshift ($z < 0.016$), hydrogen-rich Type~IIP supernovae
(SNe~IIP). 
Our newly constructed sample contains 91 nebular spectra of 38 
SNe~IIP, which is the largest dataset of its kind ever analysed in one
study, and many of the objects have complementary photometric data. 
We determined the peak luminosity, total luminosity, velocity of the
peak, half-width at half-maximum intensity, and profile shape for many permitted 
and forbidden emission lines. Temporal evolution of these values,
along with various flux ratios, are studied and compared to previous
work. We also investigate the correlations between these measurements
and photometric observables, such as the peak and plateau absolute
magnitudes and the late-time light curve decline rates in various
optical bands. The strongest and most robust result we find is that
the luminosities of all spectral features (except those of helium) tend
to be higher in objects with steeper late-time $V$-band decline
rates. A steep late-time $V$-band slope likely arises from less
efficient trapping of $\gamma$-rays and positrons, which could be
caused by multidimensional effects such as clumping of the ejecta or 
asphericity of the explosion itself. Furthermore, if $\gamma$-rays and 
positrons can escape more easily, then so can photons via the observed
emission lines, leading to more luminous spectral features. 
It is also shown that SNe~IIP with larger progenitor stars have ejecta
with a more physically extended oxygen layer that is well-mixed with
the hydrogen layer. In addition, we find a subset of objects with
evidence for asymmetric $^{56}$Ni ejection, likely bipolar in
shape. We also compare our observations to theoretical late-time
spectral models of SNe~IIP from two separate groups and find
moderate-to-good agreement with both sets of models. Our SNe~IIP
spectra are consistent with models of 12--15~\msun\ progenitor stars
having relatively low metallicity ($Z \le 0.01$). 
\end{abstract}



\section{Introduction}\label{s:intro}

Supernovae (SNe) provide a driving force for the chemical evolution of
galaxies. They return material to the interstellar and intergalactic
medium to be recycled and used in galaxy and star formation, and they
produce some of the heaviest naturally occurring elements. Type II SNe
(SNe~II) result from the core collapse of a massive ($\ga$8~\msun), 
hydrogen-rich star that has produced an iron core at the end of its
life. The collapse sends a shock wave through the stellar material
that disrupts the star. While stellar physics implies that typical
SNe~II could come from stars with masses up to 30~\msun, direct
observations have yielded progenitors with masses of only 8--16~\msun;
this is known as the red supergiant (RSG) problem
\citep{Smartt09}. Possibly related to this mystery are the recent
results that \about19~per~cent of massive, apparently single stars are
in fact the result of a merger \citep{deMink14} and that very few
model stars over 20~\msun\ explode as Type IIP SNe
\citep[SNe~IIP; e.g.,][]{Sukhbold16}.   

SNe~IIP are classified by their spectra, which are
dominated in their photospheric phase by P Cygni profiles of H Balmer
lines \citep[e.g.,][]{Barbon79,Filippenko97}, and by their light curves,
which have a 80--120~d plateau in the $R$ and $I$ bands
\citep[e.g.,][]{Faran14}, from which the ``P'' in SNe~IIP comes. Their
progenitor  
stars have a thick hydrogen envelope at the time of explosion,
leading to this signature plateau in the light curve. The plateau phase
ends once all of the hydrogen in the envelope has recombined. At
this point SN~IIP light curves show a rapid and steep drop (1--2~mag) 
before settling onto a linear decline in magnitude space
\citep[e.g.,][]{Faran14}. The energy source at these late times is the
deposition of $\gamma$-rays and positrons that come from the decay
chain $^{56}$Ni$\rightarrow$$^{56}$Co$\rightarrow$$^{56}$Fe, with most of the
energy at these epochs coming from the second step of this
process. This final phase in the life of a SN~IIP is known as
the radioactive tail (referring to the light-curve power source) or
the nebular phase (referring to the spectra, which consist mostly of
forbidden emission lines). We focus our efforts in this paper on the
spectral observations at such late epochs. 

About 40~per~cent of all SNe in a volume-limited sample are Type~IIP,
making them the most common SN subtype \citep{Li11a}. However, owing to
their relative faintness at late times \citep[$-13$ to $-16$~mag
during the nebular phase; e.g.,][]{Faran14}, it is difficult to study
SNe~IIP at these epochs. Of order two dozen previous studies of
individual SNe~IIP have included nebular spectra, while only a few
published works have presented late-time SN~IIP spectra of many
objects at once
\citep{Turatto93,Maguire12,Spiro14,Valenti16}. One of the
largest and most comprehensive studies, \citet{Maguire12} analysed
35 late-time spectra of 9 SNe~IIP. The sample studied herein consists
of 91 nebular spectra of 
38 SNe~IIP, making it the largest dataset of late-time SN~IIP spectra
ever analysed in a single study.

The data used in this work are summarised in Section~\ref{s:data}, and
our methods of nebular-phase determination and spectral-feature
measurement, as well as our late-time photometry calculations, are
described in Section~\ref{s:methods}. The analysis of our 
spectral measurements and their possible correlations with each other
and other SN~IIP observables can be found in Section~\ref{s:analysis},
while a comparison of our spectral data to theoretical models is 
discussed in Section~\ref{s:models}. We summarise our conclusions in 
Section~\ref{s:conclusions}.


\section{Dataset}\label{s:data}

\subsection{Spectroscopy}

To compile the late-time SN~IIP spectral dataset used herein, we first
searched the UC Berkeley Filippenko Group's SuperNova Database
\citep[SNDB;][]{Silverman12:BSNIPI} for spectra obtained at least 80~d
after discovery for all objects classified as SNe~II or SNe~IIP. After an
initial visual inspection, we removed a handful of these spectra
that showed strong H Balmer absorption features, indicating that the
SN was not yet in the nebular phase. To augment this initial sample,
from 2012 through 2014 we undertook a concerted observing campaign to obtain
more late-time spectra of SNe~IIP using multiple telescopes. This
yielded an additional 20 spectra of 9 SNe~IIP, increasing our dataset
by \about50~per~cent.

About half of the spectra in the present sample were obtained using
the Kast double spectrograph \citep{Miller93} on the Shane 3~m
telescope at Lick Observatory. The rest of the spectra were obtained 
with a variety of instruments and telescopes including the UV Schmidt
spectrograph \citep{Miller87} on the Shane 3~m telescope at Lick, the
Low Resolution Imaging Spectrometer \citep[LRIS;][]{Oke95} on the 10~m
Keck telescope, the DEep Imaging Multi-Object Spectrograph
\citep[DEIMOS;][]{Faber03} also on the Keck telescope, the Marcario
Low-Resolution Spectrograph \citep[LRS;][]{Hill98} on the 9.2~m
Hobby-Eberly Telescope (HET) at McDonald Observatory, and the Wide-Field
Spectrograph \citep[WiFeS;][]{Dopita07,Dopita10} on the 2.3~m Advanced
Technology Telescope at Siding Spring Observatory. All data were
reduced using modern reduction methods; for more information
regarding the data reduction, see \citet{Silverman12:BSNIPI}.

\subsection{Photometry}

Although the present work concentrates on late-time spectroscopic
observations of SNe~IIP, complementary photometric data at both early
and late times can be
informative as well. For each object in our sample, we 
conducted a literature search for both early- and late-time
photometric data. The early-time data allow us to constrain the
dates of explosion and peak magnitude, while the late-time data can be
compared to our spectra obtained at similar epochs.

For any object where no explosion date was found, we instead use the
midpoint between the date of the last nondetection and the date of
discovery. Our 
uncertainty on this date is then half the time between the date of
discovery and the date of the last nondetection. If, however, the
time between discovery and the last nondetection is $>$40~d, we
instead define the explosion date as the date of discovery minus 20~d
with an uncertainty of 20~d.

For SNe~IIP, the magnitude at discovery is often reported in the
IAU Circulars and ATels, as opposed to the peak magnitude or the magnitude on the
plateau. We assume this ``discovery magnitude'' is effectively equal
to the plateau magnitude since most SNe~IIP are discovered while in
the plateau phase. This assumption is supported by the fact that in
many cases observations of the same SN~IIP reported soon after
discovery and within a few days of each other yield consistent
magnitudes. Sometimes $R$-band or unfiltered magnitudes are reported,
but we ignore these in favour of the more often used $V$-band
observations for the peak and plateau magnitudes.

\subsection{Possible SN~IIL Contamination}

When constructing a dataset of SN~IIP observations, one needs to be
aware of possible contamination from Type IIL supernovae (SNe~IIL),
which have a ``linear'' (in mag per day) decline in their light curves instead of a
plateau \citep[e.g.,][]{Barbon79}. The long-standing distinction
between SNe~IIP and SNe~IIL has recently been scrutinised as 
progressively larger samples of well-sampled SN~II light curves have become
available. While some authors support this separation into two
separate subclasses, others argue that there exists a continuum of
light-curve decline rates
\citep[e.g.,][]{Arcavi12,Anderson14b,Faran14b,Faran14,Sanders15}. Using
the findings of these previous studies, we searched the spectra and
light curves of all objects in our sample for characteristics of
SNe~IIL. This led to a small number of objects being
removed.\footnote{e.g., SNe~2000dc, 2001cy, 2001fa, and 2008es.} 

\subsection{The Final Sample}\label{ss:final_sample}

After further analysis to determine whether a spectrum was truly
nebular (see Section~\ref{ss:old} below for further details), our
final sample consists of 91 late-time spectra of 38 SNe~IIP, 21 of
which have multiple spectra in the dataset. More than half of the
SNe~IIP in our sample (21) have never been studied previously at late
times and nearly three-quarters (66) of the spectra analysed herein are
previously unpublished. Most objects in the sample were discovered by
``targeted'' surveys which favour more-luminous host galaxies, with 34
out of the 38 SNe~IIP coming from NGC galaxies. 
Information regarding each object and spectrum
in our dataset can be found in Tables~\ref{t:objects} and
\ref{t:spec}, respectively. Upon acceptance of this paper, all 
spectra herein
will be available in electronic format in the
SNDB,\footnote{http://heracles.astro.berkeley.edu/sndb} WISeREP
\citep[the Weizmann Interactive Supernova data
REPository;][]{Yaron12},\footnote{http://www.weizmann.ac.il/astrophysics/wiserep}
and the Open Supernova Catalog
\citep{Guillochon16}.\footnote{https://sne.space} 

\begin{figure*}
\centering
\includegraphics[width=7.1in]{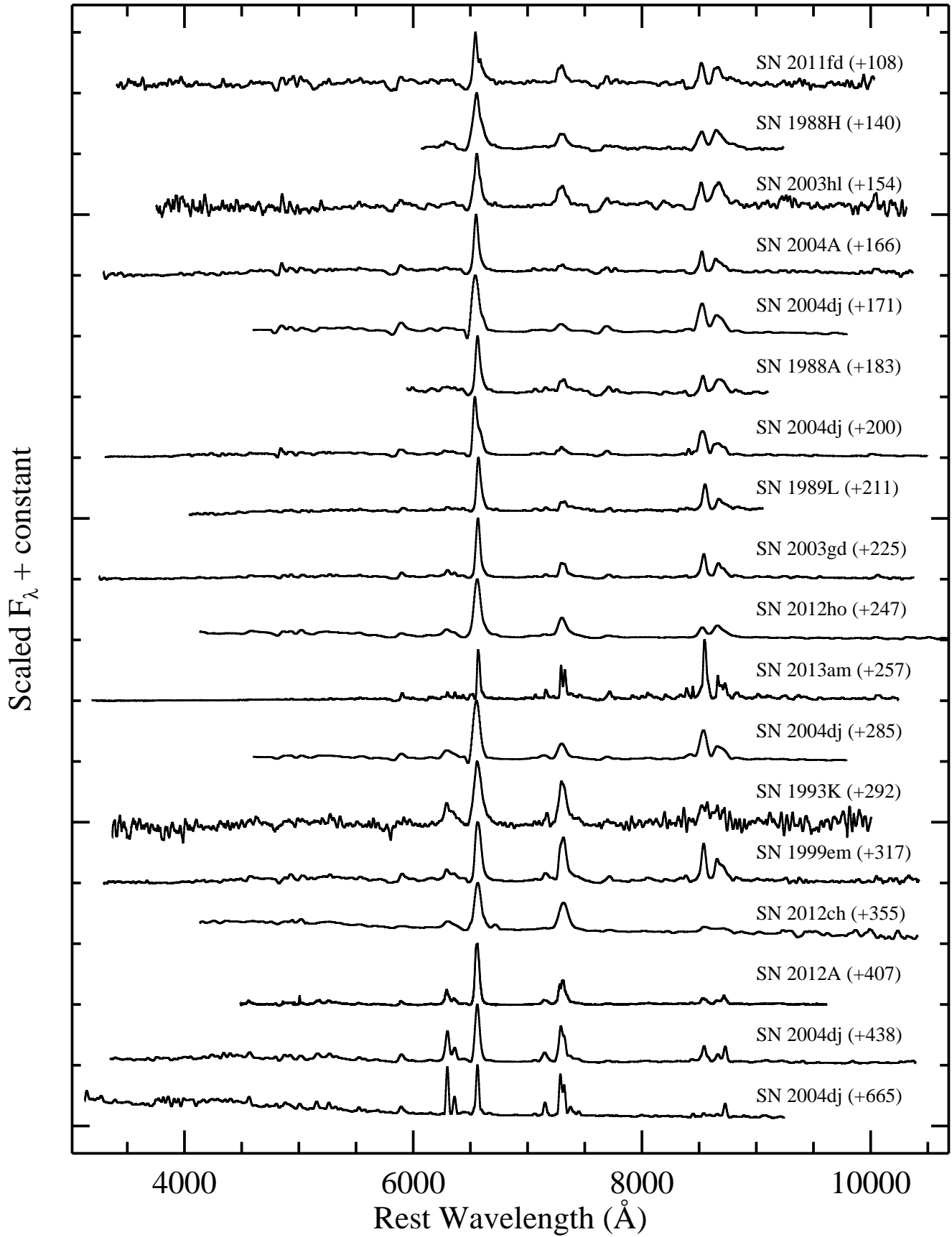}
\caption{A subset of spectra from our final sample. Each spectrum is
  labeled with the object name and its rest-frame age relative to
  explosion. All data have 
  been corrected for host-galaxy recession velocity and Galactic
  reddening using the values listed in
  Table~\ref{t:objects}.}\label{f:many_spec} 
\end{figure*}

The rest-frame ages, with respect to explosion, of the spectra studied
in this work range from 103~d to 1229~d. While \about100~d may seem
young for a SN~IIP to be in its nebular phase, our analysis in
Section~\ref{ss:old} indicates that this can indeed happen. Owing to
their relative faintness, it is unsurprising that our sample consists
of only very low-redshift objects ($z \la 0.016$) with a typical
redshift of about 0.005. Figure~\ref{f:many_spec} shows a subset of
the spectra in our final sample from a variety of epochs.

All 38 objects in our sample have published $V$-band absolute plateau
magnitudes (\mplat), the typical value of which is about
$-16.3$~mag, consistent with previous work on SN~IIP light curves
\citep[e.g.,][]{Li11a}. For nearly two-thirds (25) of the SNe~IIP in our
dataset we measure $V$-band absolute peak magnitudes (\mpk) from their
light curves, and the typical value of this is approximately
$-16.6$~mag. Late-time photometry was acquired for a different subset
of 25 objects in our sample, often in multiple bands. More on the
late-time light curves can be found in Section~\ref{ss:phot}.

\section{Methods}\label{s:methods}

\subsection{How old is old?}\label{ss:old}

To arrive at our final sample of 91 spectra of 38 SNe~IIP
(Section~\ref{ss:final_sample}), we first needed to determine whether
each SN~IIP 
spectrum was truly nebular, or, equivalently, whether the
SN~IIP was on the radioactive tail at the epoch when a given spectrum
was obtained. Previous work on SN~IIP observations often defined the 
late-time nebular phase to begin after some specified epoch, ranging
from \about150 to \about250~d past explosion; for example,
\citet{Elmhamdi03a} use 200~d after explosion. Therefore, our first
step was to calculate the rest-frame age of each spectrum in our
possible sample of 117 spectra of 55 SNe~IIP (mostly from the SNDB)
with respect to both date of explosion and date of discovery.

Upon visual inspection of these data, however, we found that some
relatively young spectra (\about100~d after explosion) appeared to be
nebular (i.e., nearly no continuum emission or absorption
features). On the other hand, some spectra from significantly later
epochs (\about170~d past explosion) were clearly not yet in the
nebular phase. Thus, it seems unwise to define the beginning of
the nebular phase of all SNe~IIP to be a semi-arbitrarily chosen
epoch. Different objects having different ages at which they
transition to the nebular phase is unsurprising given that SNe~IIP
evolve at different rates and have a range of plateau lengths
\citep[e.g.,][]{Faran14}. We therefore employ a less strict, but more
robust, determination of whether a given spectrum is truly nebular.

As mentioned above, our first method of removing spectra that were
obviously not nebular was a visual inspection and
comparison to high signal-to-noise ratio (S/N) spectra of very
late-time SN~IIP data. We then investigated whether the \hal\ 
profiles, the strongest feature in SN~IIP spectra, showed evidence for
P Cygni absorption. If present, this would be an indicator
of an optically thick photosphere which should not exist in nebular
phases. Unfortunately, our spectra often did not have sufficiently high 
S/N to confidently determine whether \hal\ absorption was present. 

Our main method of identifying nebular-phase spectra involved
forbidden emission lines of oxygen ([\ion{O}{I}]
$\lambda\lambda$6300, 6364) and calcium ([\ion{Ca}{II}]
$\lambda\lambda$7291, 7324). At our desired late epochs, SN~IIP
spectra are dominated by emission features, including many
forbidden lines, so the presence of [\ion{O}{I}] and [\ion{Ca}{II}]
should be a reasonable indicator of being in the nebular phase. Using
our spectral feature fitting routine (see Section~\ref{s:measuring}
for further details), we attempted to fit a double-Gaussian function to
[\ion{O}{I}] $\lambda\lambda$6300, 6364 and [\ion{Ca}{II}]
$\lambda\lambda$7291, 7324 in each spectrum. The feature was considered
to be present if the peaks of the Gaussian fits were $>2\sigma$
above the locally determined continuum and two distinct peaks were
detected. [\ion{O}{I}] $\lambda\lambda$6300, 6364 is often the stronger
of the two features investigated, so this was the main indicator of
whether a spectrum was nebular, though [\ion{Ca}{II}]
$\lambda\lambda$7291, 7324 was also detected in many of the same
spectra. This search for forbidden lines mostly supported our visual
inspection and comparison to high-S/N nebular spectra mentioned
above. 

Furthermore, a majority of the SNe~IIP we investigated had
companion photometry (see Section~\ref{ss:phot} for more
information). Using these data we were able to photometrically
determine when many of our objects entered the radioactive tail phase
of their light curve. For each SN~IIP with photometric data, the
epochs at which our nebular spectra were obtained were all found to be
after the object began its late-time photometric decline. Thus, our spectroscopic
and photometric determinations of the beginning of the nebular phase
are consistent. This analysis resulted in narrowing the sample from
117 spectra of 55 SNe~IIP to our final sample of 91 spectra of 38
SNe~IIP (Section~\ref{ss:final_sample}).

\subsection{Measuring Nebular Spectral Features}\label{s:measuring}

The routine used in this work to measure the emission features in the
late-time spectra of SNe~IIP is similar to that used to measure
emission features in SN~Ia spectra also obtained from the SNDB 
\citep{Silverman12:BSNIPII,Silverman13:late}. The method is described 
in detail in previous work, but here we give a brief summary of the
procedure. 

Each spectrum is first corrected for its host-galaxy
recession velocity and Galactic reddening using the values listed in
Table~\ref{t:objects}, then smoothed using a Savitzky-Golay smoothing
filter \citep{Savitzky64}. Reddening from the host galaxies is not
removed, as most of the objects in the current sample appear to
be relatively unreddened by their hosts (i.e., they lack obvious
narrow \ion{Na}{I}~D absorption in our spectra and in publicly
available early-time spectra). The significant exception to this
statement is SN~2002hh, which has \about6~mag of extinction from its
host \citep{Welch07}.

Since all of the spectra in our sample are nebular, the continuum
level should be nearly nonexistent, so we do not include any
background or continuum level in our fits. For each feature
investigated, the endpoints of the emission profile are chosen by hand
and the data between these endpoints are then fit with a cubic spline
as well as a (multi-)Gaussian function. The number of Gaussians used in
each fit depends on the number of detectable, but blended, features in
the profile. 


It was found that the cubic spline fits captured the peaks of each
spectral feature much more accurately than the Gaussian
fits. Therefore, the peak of each spline fit was recorded as the peak
flux (\fpk) and the wavelength at which this peak occurred was used to
calculate the peak velocity (\vpk) using the relativistic Doppler
equation. We also recorded the total flux (\ftot) in each spectral
feature between the aforementioned endpoints. When comparing multiple
SNe, it is more instructive to use luminosities than observed fluxes,
so we converted all of our measured \fpk\ and \ftot\ values to
luminosities (\lpk\ and \ltot, respectively). This was accomplished by
using the mean metric distance to each SN~IIP (from NED and listed in
Table~\ref{t:objects}) or, for the four objects without measured
distances, using the redshift (also from NED and listed in
Table~\ref{t:objects}) 
and $H_0 = 73$~km~s$^{-1}$~Mpc$^{-1}$ \citep[][]{Riess16}. The values
of \lpk, \ltot, and \vpk\ for each spectral feature can be found in
Tables~\ref{t:hydrogen}--\ref{t:iron}.

On the other hand, the Gaussian fits appeared to capture the widths
of each spectral feature better than the spline fits. Thus, the
Gaussian fits were used to determine the half-width at half-maximum intensity
\citep[HWHM; e.g.,][]{Maguire12} and the half-width at zero intensity
(HWZI) of each feature. As in \citet{Maguire12}, both of these
parameters were corrected for the instrumental resolution of each
spectrum (listed in Table~\ref{t:spec}) using an equation of the form
$${\rm HWHM}_\textrm{corrected} = \sqrt{{\rm HWHM}_\textrm{measured}^2 - 
{\rm HWHM}_\textrm{resolution}^2}.$$
When comparing the measured values of the HWHM and HWZI, it was found
that in the vast majority of cases HWZI~$\approx 2.36 \times$HWHM as a result
of our Gaussian fitting method. Thus, our HWZI measurements did not
yield any new information and are ignored throughout the rest of this
work. The HWHM for each spectral feature is listed in
Tables~\ref{t:hydrogen}--\ref{t:iron}.

In addition to the cubic spline and Gaussian fits, each spectral
feature was assigned a descriptor of its overall visual
shape or appearance: single-peaked, multi-peaked, flat-topped, or
other. For extremely low-resolution spectra, the appearance of the
spectral profile may not reflect the actual underlying profile, but
\citet{Maguire12} found that for resolutions of \about14~\AA\ or better
one can distinguish the true shape of moderately strong emission
features. Some of the spectra in our dataset are near this cutoff, but
the vast majority have higher resolution. Thus, we should be able to
determine the intrinsic shape of the stronger emission profiles in our
spectra. In short, we do not detect any flat-topped profiles, though
we do find a handful of asymmetric, double-peaked
profiles. Section~\ref{ss:shape} discusses the profile shapes in our
sample in much greater detail. 

\begin{figure*}
\centering
\includegraphics[width=7.3in]{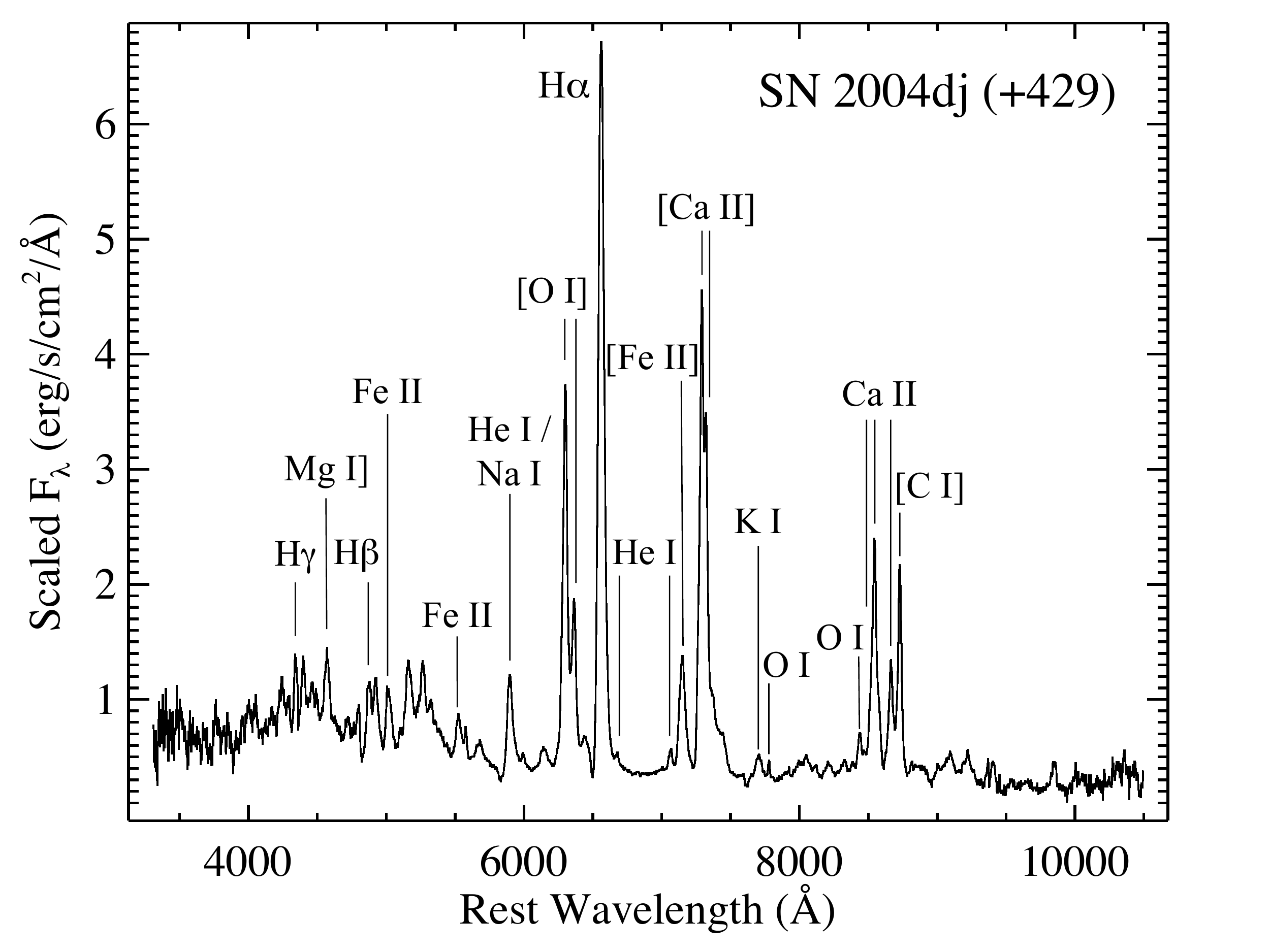}
\caption{Spectrum of SN~2004dj from 429~d past explosion with all
  features investigated herein labeled. Also labeled is [\ion{C}{I}]
  $\lambda$8727 which is clearly detected here, but rarely seen in
  most of the spectra in our sample. The spectrum has been
  corrected for its host-galaxy recession velocity and Galactic
  reddening using the values listed in
  Table~\ref{t:objects}.}\label{f:one_spec}
\end{figure*}

\begin{figure}
\centering
\includegraphics[width=3.5in]{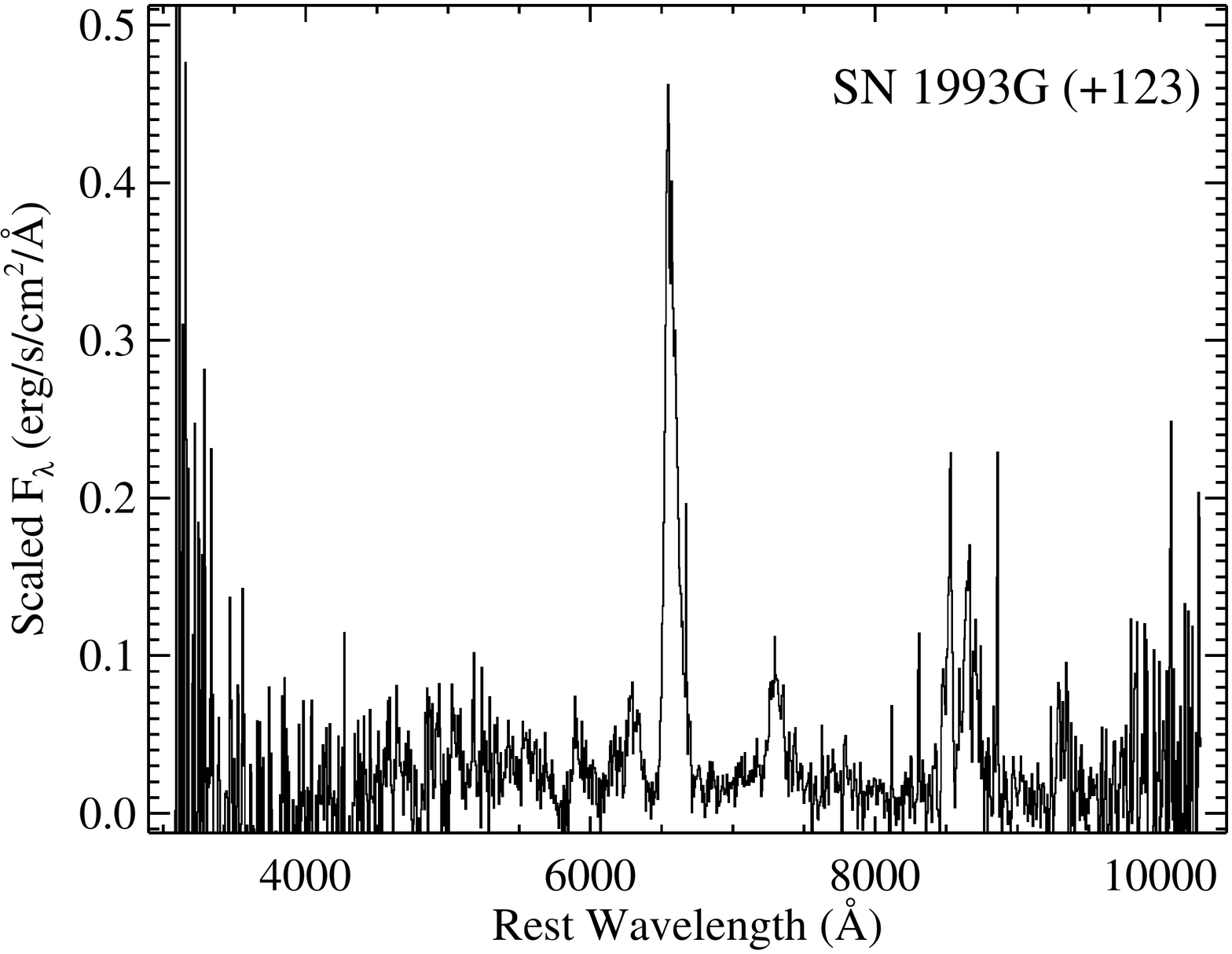}
\caption{A relatively low-S/N spectrum of SN~1993G from 123~d past
  explosion. The spectrum has been 
  corrected for its host-galaxy recession velocity and Galactic
  reddening using the values listed in
  Table~\ref{t:objects}.}\label{f:one_spec_noisy}
\end{figure}

Using the measurement technique described above, we attempt to fit
numerous features in each of our SN~IIP spectra. These consist of a
combination of permitted and forbidden emission lines of hydrogen,
helium, oxygen, magnesium, potassium, calcium, and iron. Spectral
features of other elements were considered, including carbon, sodium,
and nickel, but no or very few significant detections of these were
found in our spectra. Figure~\ref{f:one_spec} shows one of our 
highest-S/N spectra, an observation of SN~2004dj from 429~d after explosion;
all features investigated herein are labeled. On the other hand,
Figure~\ref{f:one_spec_noisy} shows one of our younger and lower-S/N
spectra, an observation of SN~1993G from 123~d past explosion. Despite
the significant noise in the spectrum, many emission features are
readily identified.

\subsection{Spectral Features Investigated}

The most prominent feature in late-time spectra of SNe~IIP is \hal\
and we clearly detect it in all 90 (out of 91 total) spectra that
extend above 6500~\AA. Significantly weaker, but still fit in most of
our spectral sample, is H$\beta$, which occasionally appears blended
with another feature near 4910~\AA. The H$\gamma$ line is even weaker, 
but still confidently detected in many of our spectra. The
measured values from the cubic spline and Gaussian fits for all
hydrogen lines can be found in Table~\ref{t:hydrogen}.

At early times, relatively strong helium lines are common in optical
spectra of SNe~IIP, but in observations at later times they are often
not as obvious \citep[e.g.,][]{Filippenko97}. Theoretical models of
nebular spectra of SNe~IIP, however, sometimes predict quite strong
helium lines \citep[e.g.,][]{Dessart13}. The strongest helium feature
in the optical is \ion{He}{I} $\lambda$5876, but the \ion{Na}{I}~D
feature is almost coincident with it and is thought to dominate the
emission profile at late times \citep{Leonard02}. Despite this, we
detect two other helium lines (\ion{He}{I} $\lambda$6678 and
\ion{He}{I} $\lambda$7065) in many of our spectra with profile shapes
consistent with those of \ion{He}{I} $\lambda$5876. The two redder
lines are more prominent in the later epochs studied herein, with
\ion{He}{I} $\lambda$7065 appearing more often than \ion{He}{I} $\lambda$6678. 
Given the consistency of these detections and the relatively low resolution
of most of our dataset, it seems likely that the often-detected
emission near \about5900~\AA\ is dominated by \ion{He}{I}
$\lambda$5876. Detailed spectral modeling could be used to determine
more precisely the relative contributions of \ion{He}{I} $\lambda$5876
and \ion{Na}{I}~D to the emission profile in each spectrum, but this
is beyond the scope of this paper. The measured values from the cubic
spline and Gaussian fits for all helium features can be found in
Table~\ref{t:helium}.

As mentioned in Section~\ref{ss:old}, the main criterion for
including a spectrum in the current dataset was the detection of the
[\ion{O}{I}] $\lambda\lambda$6300, 6364 doublet. Thus, it is not
surprising that this feature was successfully fit in nearly all (89
out of 91) of our spectra. The two observations where it was not
detected both show strong [\ion{Ca}{II}] $\lambda\lambda$7291, 7324
emission, which led to their inclusion in our sample. Owing to the
relatively low resolution of most of the spectra studied herein, the
[\ion{O}{I}] $\lambda\lambda$6300, 6364 doublet often appears somewhat 
blended, though two distinct peaks are always discernible. Because of
this, we fit this doublet with a cubic spline as before, as well as 
with a double-Gaussian function. The highest peak of the doublet, used
to calculate \fpk\ and \vpk, was always the bluer component, and \ftot\
represents the total flux in the doublet (i.e., not either individual
component). Since the Gaussian fits are used to calculate the HWHM
and a double-Gaussian function is used for this doublet, a HWHM value is
calculated for each component individually.

Two other oxygen lines are also investigated: \ion{O}{I}
$\lambda$7774 and \ion{O}{I} $\lambda$8446. Both of these are actually
triplets, but their components are so closely spaced that we cannot
resolve them and thus fit each feature as a single spectral line. In
addition, we find that \ion{O}{I} $\lambda$7774 is usually blended
with a doublet, \ion{K}{I} $\lambda\lambda$7665, 7699,
which we also fit as a single spectral feature (using the average
wavelength of the two components, 7682~\AA, as its rest
wavelength). This resonance line has been 
observed in previously published late-time spectra of SNe~IIP
\citep{Chornock10} and theoretical models support this spectral
identification \citep{Dessart13}. Thus, we fit \ion{K}{I}
$\lambda$7682 and \ion{O}{I} $\lambda$7774 as a doublet (as was
done for [\ion{O}{I}] $\lambda\lambda$6300, 6364). The peak of the
\ion{K}{I} emission was usually stronger than that of \ion{O}{I}
$\lambda$7774, so \fpk\ and \vpk\ were calculated with respect to this
feature. The measured values from the cubic spline and Gaussian fits for all
oxygen and potassium features can be found in Table~\ref{t:oxygen}.  

Only one magnesium feature is confidently identified in many of our
spectra: \ion{Mg}{I}] $\lambda$4571. It is often relatively weak,
especially in spectra from the earlier epochs in our sample, but we
are still able to fit its profile in \about60~per~cent of our
dataset. The measured values from the cubic spline and Gaussian fits for
this magnesium feature can be found in Table~\ref{t:magnesium}.

A secondary criterion for inclusion in our sample, as mentioned in
Section~\ref{ss:old}, was the detection of the [\ion{Ca}{II}]
$\lambda\lambda$7291, 7324 doublet. This relatively strong feature was
detected in almost all of our spectra, though it was sometimes blended
on its blue side with other (possibly iron) emission lines. This
feature was fit as a doublet, just like [\ion{O}{I}]
$\lambda\lambda$6300, 6364 discussed above, and \fpk\ and \vpk\ were
calculated with respect to whichever of the two components was
stronger. The bluer component had larger peak flux in about 3/4 of the 
spectra, but both components were often of very similar strength. We also
identified the \ion{Ca}{II} near-infrared (NIR) triplet
$\lambda\lambda$8498, 8542, 8662 in most of the spectra in our
sample. The bluer two components were often blended and we fit them
as a doublet, with the 8542~\AA\ feature being the stronger of the two;
thus, \fpk\ and \vpk\ values were measured with respect to that
line. The reddest feature in the triplet would sometimes be blended
with weak, but noticeable, emission on its red wing (possibly from
[\ion{C}{I}] $\lambda$8727), but this very rarely affected the fitting of the
\ion{Ca}{II} profile. The measured values from the cubic spline and Gaussian
fits for all calcium features can be found in Table~\ref{t:calcium}.

There are numerous iron emission lines (both permitted and forbidden)
from various multiplets that fall in the optical range, and this leads
to severe blending of the vast majority of these features. In spite of
this, we are able to securely and consistently identify three iron
features. Emission from both \ion{Fe}{II} $\lambda$5018 and \ion{Fe}{II}
$\lambda$5527 is detected in many of our spectra, though the lines
are sometimes relatively weak and both occasionally have blended emission
on their red wings. The strongly blended [\ion{Fe}{II}]
$\lambda\lambda$7155, 7172 doublet is also often seen in the spectra 
in our dataset and usually appears relatively broad, but with only one
distinct peak. Because of this, we fit this doublet with a
single-Gaussian function (in addition to the cubic spline), and
\vpk\ is calculated relative to 7155~\AA, the stronger of the two
blended components. We searched for other iron features, 
including [\ion{Fe}{II}] $\lambda$5164, \ion{Fe}{II} $\lambda$5169,
and \ion{Fe}{II} $\lambda$5270, but very few significant detections
were made in our spectral sample. The measured values from the cubic
spline and Gaussian fits for all iron features can be found in 
Table~\ref{t:iron}.

\subsection{Late-Time Photometry}\label{ss:phot}

While our spectra are spectrophotometrically accurate in a relative
sense, they may not be in an absolute sense owing to factors such as
slit losses or cloud cover \citep[see][for further
details]{Silverman12:BSNIPI}. This limits our ability to measure
reliable line fluxes (and luminosities), so a literature search was
conducted in order 
to gather as much accompanying optical photometry as possible for the
SNe~IIP in our sample. This photometry also allows us to compare our
late-time spectral measurements to photometric 
observables (as described in Section~\ref{s:analysis}). The
literature search yielded 77 optical ($B$, $V$, $R$, $I$, and
unfiltered) light curves of 25 objects. 15 of these light curves
(representing 7 SNe~IIP) are previously unpublished, but the
observations and data reduction pipeline are described by
\citet{Ganeshalingam10}. 

Using the 22 $V$-band light curves obtained, we measured the peak
absolute magnitude (\mpk) and the median absolute magnitude during the
plateau phase (\mplat), where the beginning and end of the plateau was
chosen manually for each light curve. For objects with no available
$V$-band photometry, \mpk\ (when available) and \mplat\ values from
the literature were used. These values are listed in Columns 8 and
9 of Table~\ref{t:objects}. 

\begin{figure*}
\centering
\includegraphics[width=7in]{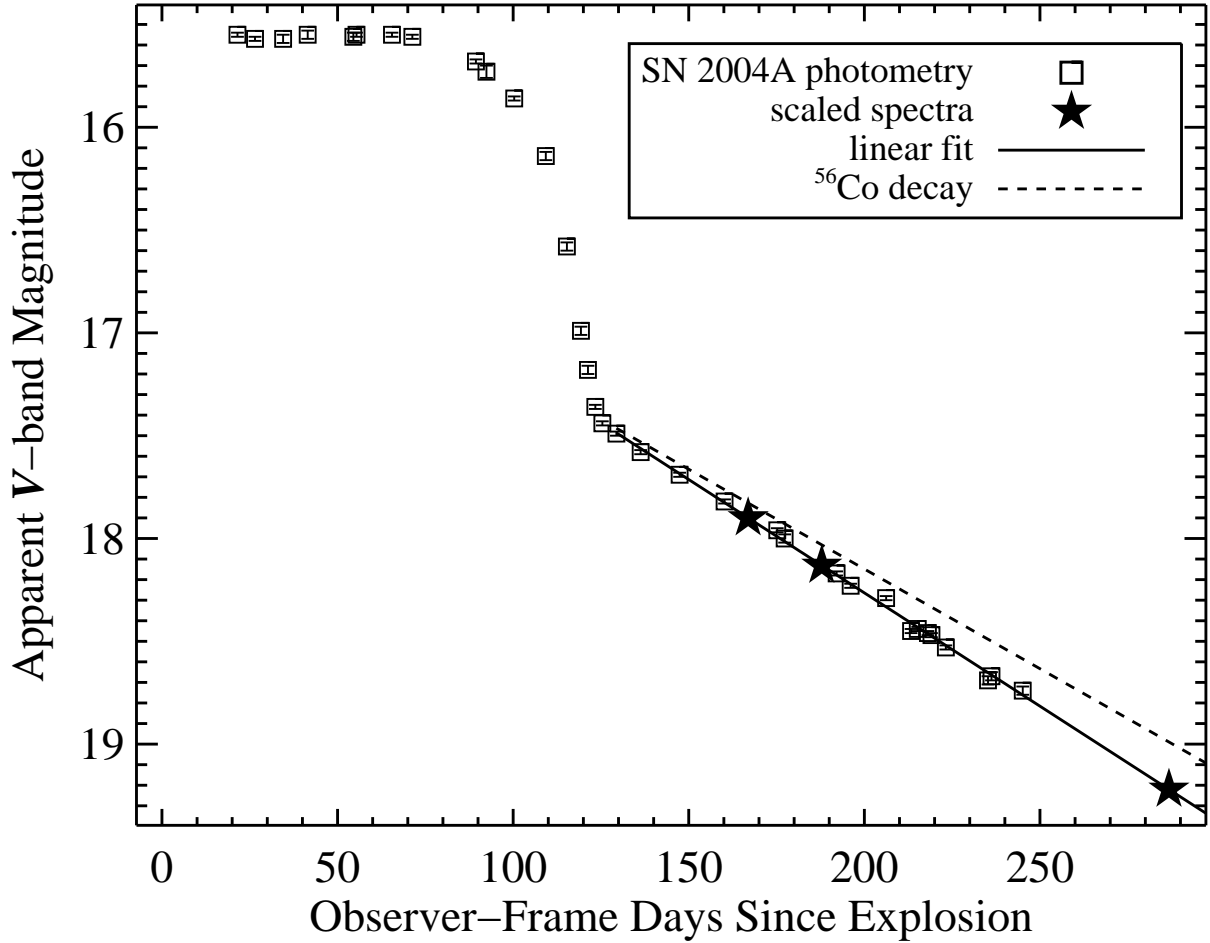}
\caption{The $V$-band light curve of SN~2004A (open squares) from
  \citet{Gurugubelli08}. Overplotted is our linear fit to the
  late-time radioactive tail (solid line) and the decay rate of
  radioactive $^{56}$Co (dashed line), which is
  0.97~mag~(100~d)$^{-1}$. The interpolated and extrapolated 
  magnitudes for days on which we have spectra of SN~2004A are also
  plotted (filled stars).}\label{f:one_lc} 
\end{figure*}

For each of the 77 light curves acquired, we also manually determined when
the late-time radioactive tail began and then fit a line to those data
points. For light curves with relatively sparse observations, it was
sometimes difficult to determine the exact beginning of the
radioactive tail, but oftentimes there was a clear separation in
magnitude-space between the plateau phase and the nebular
phase. Figure~\ref{f:one_lc} presents an example light curve
\citep[the $V$-band data of SN~2004A from][open
squares]{Gurugubelli08} along with our linear fit to the late-time
radioactive tail (solid line). Overplotted is the decay rate of
radioactive $^{56}$Co (dashed line), which is 0.97~mag~(100~d)$^{-1}$. 
Lastly, we constructed pseudobolometric late-time light curves for
all objects with photometry in at least three bands. This yielded 13
\bvri light curves, 2 \protect\hbox{$BV\!R$}\ light curves, and 4
\protect\hbox{$V\!RI$} light curves.

The results of our linear fits to the radioactive tails of the light
curves are displayed in Table~\ref{t:lc}. There we list the number of
photometric points used in each linear fit, the MJD range spanned by
those points, and the resulting slope of the linear fit and its
uncertainty. The late-time rate of decline [in mag~(100~d)$^{-1}$] of the
SNe~IIP is the main parameter in which we are interested, which is why
the slopes are the only values from the linear fits listed in the
Table. 

For a given SN~IIP, the late-time slope is steeper as the observed
bandpass gets redder, consistent with previous work
\citep[e.g.,][]{Dhungana16}. As is the case for individual SNe~IIP,
the mean and median slope for all objects in each band is largest in
the reddest bands and smallest in the bluest bands, with $B$, $V$, and
$R$ slower-declining than $^{56}$Co and $I$ declining slightly
faster than $^{56}$Co. However, given the relatively large standard deviations, the
mean slope of each of the four bands is formally consistent with the $^{56}$Co
decay rate.

That being said, we find a large range of values for the
late-time decline rates. Some of the measured slopes are much smaller
(i.e., slower or shallower) than the $^{56}$Co decay rate and some much
larger (i.e., steeper- or faster-declining). While there are no objects
with extremely steep slopes, there are three SNe~IIP with slopes that are
significantly shallower than the $^{56}$Co decay rate (by at least a
factor of two): SNe~2005cs, 2006ov, and 2013am. If our determination of
the beginning of the late-time radioactive tail was incorrect, then we
might have included data from the steep drop-off phase of the light
curve. This would then lead to a slope that is much steeper than the
$^{56}$Co decay rate, and we have no examples of such slopes, so this
supports our measurements of the beginning of the late-time tail. As
for the three objects with extremely shallow slopes, they also
exhibit some of the lowest HWHM values (i.e., narrowest emission lines)
in our sample and tend to be low-luminosity SNe~IIP, which confirms
previous work on these objects
\citep{Pastorello09b,Spiro14,Zhang14}.

As stated above, the initial impetus for gathering these photometric
data was to place our spectra on an accurate absolute
spectrophotometric scale. To do this, we follow a procedure similar
to what was used for SNe~Ia spectra by \citet{Silverman12:BSNIPI}. We
first calculated synthetic magnitudes from each spectrum of all
objects where we were able to obtain late-time photometry. We then
interpolated or extrapolated our linear fits of the radioactive decay
phase of the light curves in order to calculate the photometric
magnitude of each object for days on which we have
spectra. Table~\ref{t:spec_phot} lists these interpolated/extrapolated
magnitudes and their uncertainties in each filter. The filled stars in
Figure~\ref{f:one_lc} represent the $V$-band magnitude of SN~2004A
for the three days on which we have spectra.

The synthetic magnitudes from a given spectrum were then compared to
the photometric magnitudes (from our linear fits to the late-time
light curves) 
on that same day in order to calculate a scale factor for each
band. The flux values of the spectrum were then multiplied by the
median of these scale factors in order to place it on an accurate
absolute flux scale. We investigated other methods of scaling the
spectral flux values, namely using only the $R$-band scale factor
(since this bandpass contains \hal, the strongest feature in each
spectrum) and using the mean of the scale factors. In both cases the
results were similar to using the median of the scale factors, and
the median led to more consistent results for all spectra of a given
SN~IIP.

\section{Analysis}\label{s:analysis}

Using the measured values displayed in
Tables~\ref{t:hydrogen}--\ref{t:iron}, we investigated the temporal
evolution of each parameter for each spectral feature. We also
compared our spectroscopic measurements to the photometric observables
\mpk, \mplat, and the late-time slope in each optical photometric band
(as described in Section~\ref{ss:phot}).

\subsection{Total Luminosity (\ltot), Peak Luminosity (\lpk)}\label{ss:ltot_lpk}

In previous work, \ltot\ of the strongest emission features (i.e.,
\hal, [\ion{O}{I}] $\lambda\lambda$6300, 6364, and [\ion{Ca}{II}]
$\lambda\lambda$7291, 7324) has been measured for a 
handful of individual SNe~IIP. The values we measure for SN~1999em,
for example, are very similar to those presented by
\citet{Elmhamdi03b} at similar epochs. On the other hand, our \ltot\
values of these features in SNe~2004et and 2012ec are somewhat lower
that what was found by \citet{Sahu06} and \citet{Jerkstrand15},
respectively.

\begin{figure}
\centering
\includegraphics[width=3.5in]{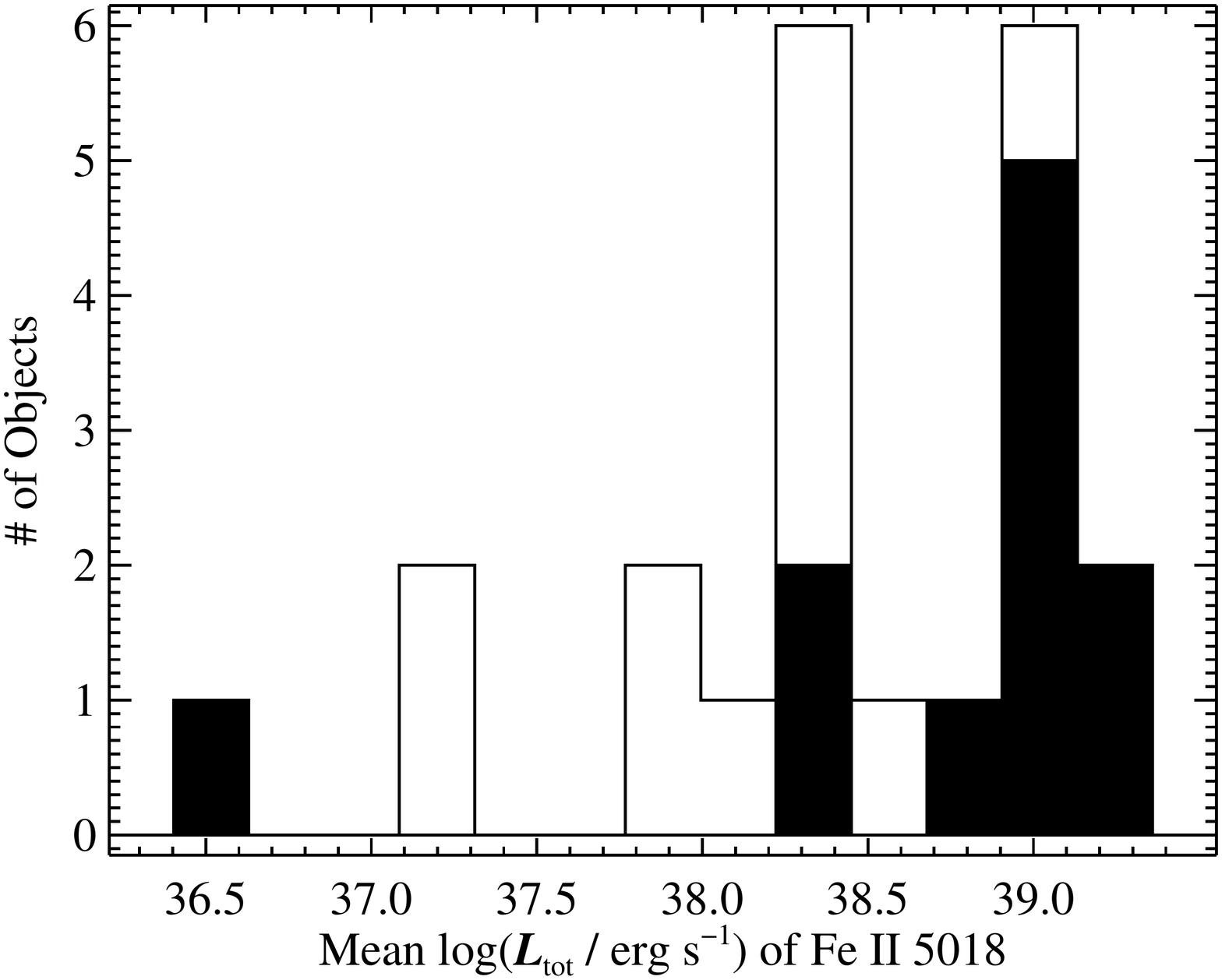} 
\caption{A histogram of the logarithm of the mean \ltot\ of
  \ion{Fe}{II} $\lambda$5018 for each object. Filled regions 
  represent objects with \mplat\ brighter than or equal to
  $-16.3$~mag; unfilled regions represent objects with \mplat\ 
  fainter than $-16.3$~mag. The objects with more luminous plateaus
  tend to have larger values of \ltot, implying a direct connection
  between the energy in the SN ejecta during the plateau and at later
  times.}\label{f:ltot_fe5018_mplat_hist}  
\end{figure}

The values of \lpk\ and \ltot\ for the bluer spectral features
investigated in this work tend to be higher for SNe~IIP with brighter
\mplat. A Kolmogorov-Smirnov (KS) test indicates that \lpk\ and \ltot\ 
values of these features for objects with brighter \mplat\ 
statistically differ from those of objects with fainter \mplat\ ($p
= 0.001$--0.04 for \lpk\ and \ltot\ with various bright/faint cutoff
values). The specific 
case of mean \ltot\ values of \ion{Fe}{II} $\lambda$5018 for each
object is shown in Figure~\ref{f:ltot_fe5018_mplat_hist}. The blue
features that exhibit this difference are included in the $B$ and $V$
bands, and it has been seen previously that a more luminous plateau
will lead to more luminous lines in these bands at late times 
\citep[e.g.,][]{Valenti16}, matching what is found herein. According
to theoretical models, \mplat\ tends to be brighter for larger 
progenitor radius and \ltot\ for all emission lines should be higher
for larger $^{56}$Ni production
\citep[e.g.,][]{Hamuy03,Spiro14,Pejcha15,Valenti16}. Thus, taking our
measured correlation a step further, one might expect a positive
correlation between progenitor radius and mass of $^{56}$Ni produced,
but models show only a moderate connection between these physical
parameters \citep{Dessart11}.

\begin{figure}
\centering
\includegraphics[width=3.5in]{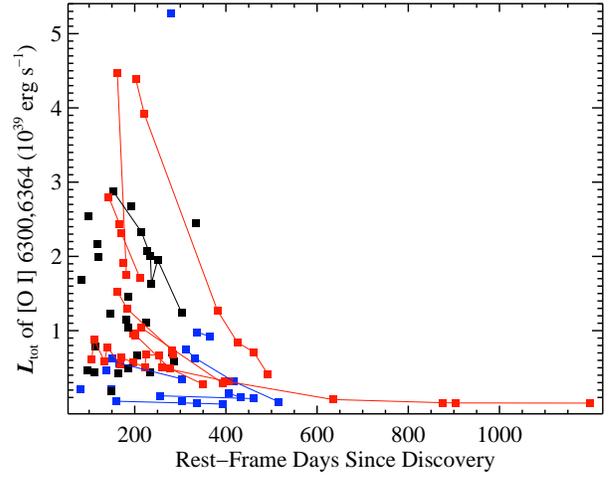} 
\caption{\ltot\ of [\ion{O}{I}] $\lambda\lambda$6300, 6364 versus
  time. Blue points are SNe~IIP with slower/shallower late-time
  decline rates than the $^{56}$Co decay rate; red points have
  faster/steeper decline rates; black points have no late-time
  $V$-band photometry. Squares represent profiles that are fit well by
  a double-Gaussian function (as this feature is a doublet); triangles
  represent more complex-shaped profiles (see Section~\ref{ss:shape}
  for more information). Filled points are spectra that have been
  scaled to contemporaneous photometry; open points have not been
  scaled. Spectra of the same object are connected with
  solid lines. Uncertainties on these \ltot\ measurements are
  typically smaller than the data points.}\label{f:ltot_oi_vslope} 
\end{figure}

One of the strongest and most robust correlations discovered in this
work is that \ltot\ and \lpk\ values for all spectral features (except
those of helium) tend to be higher for steeper late-time $V$-band
slopes. According to a KS test, \lpk\ and \ltot\ values for objects
with late-time $V$-band slopes steeper than the $^{56}$Co decay rate
are statistically different than those of objects with shallower
$V$-band slopes ($p = 0.001$--0.05 for all non-helium features studied
herein). The \ltot\ values for [\ion{O}{I}] $\lambda\lambda$6300, 6364
versus time are shown in Figure~\ref{f:ltot_oi_vslope}. Objects with
slower/shallower late-time decline rates than the $^{56}$Co decay rate
are shown in blue while objects with faster/steeper decline rates are
shown in red; black points are objects with no late-time $V$-band
photometry. Similar results are obtained when using the median
late-time $V$-band slope as the cutoff instead of the $^{56}$Co decay 
rate. The relatively few pseudobolometric late-time slopes in
our sample show the same correlation, though slightly weaker, and
there is some indication that the correlation also holds for late-time
$B$-band slopes as well. On the other hand, there is no significant
correlation between \lpk\ and \ltot\ values and late-time $R$- and
$I$-band slopes.

\begin{figure}
\centering
\includegraphics[width=3.5in]{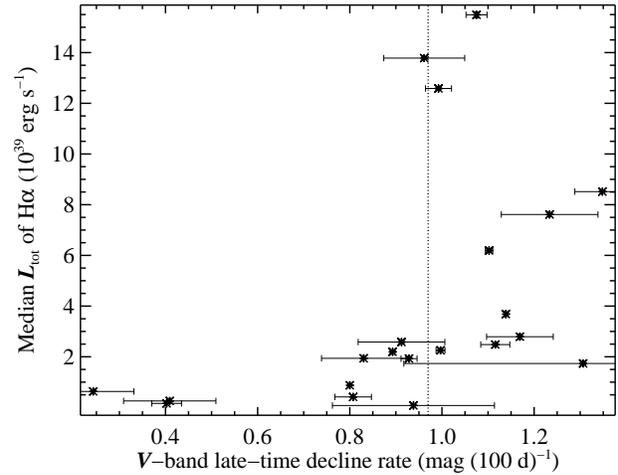}
\caption{Median \ltot\ values of \hal\ versus $V$-band late-time decline
  rate. 
  The vertical
  dotted line is the $^{56}$Co decay rate; objects to the right of
  this line (i.e., ones with faster/steeper $V$-band decline rates)
  tend to have larger values of \ltot.}\label{f:hal_vslope}
\end{figure}

To further investigate this correlation, we directly compared the
late-time $V$-band slopes with the median measurements of \lpk\ and
\ltot\ (as well as \vpk\ and HWHM) for each SN~IIP. Consistent with
our results discussed above, the $V$-band slopes were found to
correlate with \lpk\ and 
\ltot, having Pearson correlation coefficients of \about0.3--0.6, and
\ltot\ showing slightly stronger correlations than \lpk. As a specific 
example, Figure~\ref{f:hal_vslope} shows that the median values of
\ltot\ of \hal\ for the 21 SNe~IIP with late-time $V$-band photometry are
positively correlated with $V$-band slope. 
Also, objects with $V$-band decline rates that are faster/steeper than
the $^{56}$Co decay rate (i.e., to the right of the vertical dotted
line in the Figure) tend to have larger values of \ltot. These results
are effectively unchanged if we instead use the minimum, maximum,
earliest, latest, or mean values of \ltot\ and \lpk. The median HWHM
values also tend to be larger for objects with larger/steeper $V$-band
slopes, especially in the strongest emission lines (i.e., \hal,
[\ion{O}{I}] $\lambda\lambda$6300, 6364, and [\ion{Ca}{II}] 
$\lambda\lambda$7291, 7324); see Section~\ref{ss:hwhm} for more
information.

A steep late-time $V$-band slope likely arises from less efficient
trapping of $\gamma$-rays and positrons, but there are multiple
explanations for this. A relatively small hydrogen envelope may
not be sufficiently large or dense to trap $\gamma$-rays and
positrons efficiently at late time \citep[e.g.,][]{Anderson14b}. In
addition, multidimensional effects such as clumping of the ejecta or
asphericity of the explosion itself may lead to inefficient trapping
\citep{Dessart11b}. Furthermore, at least in Type~Ia SNe, the
deposition of $\gamma$-rays and positrons is likely dominated by the
strength and distribution of magnetic fields
\citep[e.g.,][]{Penney14}. Another explanation for steep late-time
decline rates, especially in the bluer optical bands, is the formation
of dust which will reprocess blue light into red/infrared light
\citep{Sahu06}. This explanation seems unlikely for the spectra
studied herein, however, since significant dust formation is thought
to begin more than \about400~d after explosion, which is significantly
older than the majority of our observations. The dust-formation
explanation is also likely ruled out by our analysis of the observed
profile shapes in Section~\ref{ss:shape}. 

On the other hand, the three objects toward the left side of
Figure~\ref{f:hal_vslope} (SNe~2005cs, 2006ov, and 2013am) have
extremely shallow late-time slopes as well as some of the narrowest
emission lines in our sample, and they also tend to be low-luminosity
SNe~IIP \citep[see also][]{Pastorello09b,Spiro14,Zhang14}. Shallow
late-time slopes have sometimes been attributed to the
presence of light echoes \citep[e.g.,][]{Otsuka12}, but this effect
also usually appears much later than nearly all of the epochs
investigated herein. Instead, there must be some other energy source
in addition to the decay of $^{56}$Co. Perhaps larger
amounts of other radioactive elements are produced in these objects as
compared to the rest of the sample, or additional radiation is being
generated in the warmer inner ejecta and propagating into the
optically thin and cooler external layers \citep{Utrobin07b}. Extra
energy could also come from the SN~IIP ejecta interacting with 
circumstellar material, but we find no spectral signatures of such
interaction in our dataset (see Section~\ref{ss:shape}). 

Regardless of the root physical cause of the steeper late-time
$V$-band decline rates, if $\gamma$-rays and positrons can more easily
leak out of the SN ejecta, then so can optical photons via the
observed emission lines. This would naturally lead to more luminous
spectral features, as we observe. Furthermore, models have shown that
more massive progenitors have stronger late-time emission features {\it
  and} smaller hydrogen envelopes, which would allow more $\gamma$-ray
and positron leakage at late times, and thus a steeper light-curve
decline \citep{Anderson14b}.

\subsection{Peak Velocity (\vpk)}\label{ss:vpk}

\begin{figure}
\centering
\includegraphics[width=3.4in]{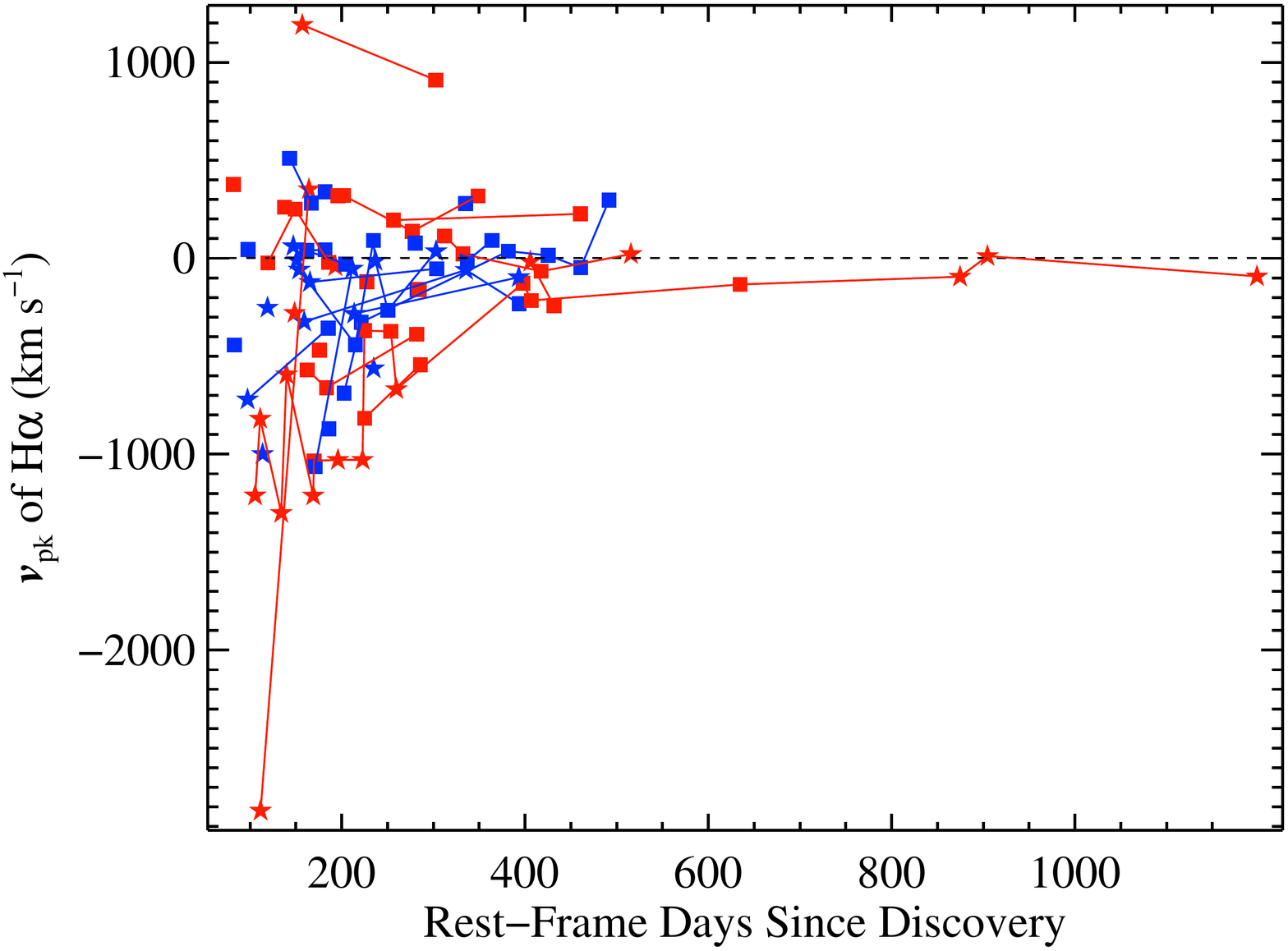} 
\caption{The \vpk\ of \hal\ versus time. Blue points are SNe~IIP with
  \mplat\ brighter than or equal to $-16.3$~mag; red points have
  \mplat\ fainter than $-16.3$~mag. Squares represent profiles that
  are fit well by a single-Gaussian function; stars represent profiles
  whose shapes are more complex (see Section~\ref{ss:shape} for
  more information). Spectra of the same object are connected with
  solid lines. Uncertainties in these \vpk\ measurements are
  typically smaller than the data points. The horizontal dashed line
  is at zero velocity.}\label{f:vpk_hal}
\end{figure}

Our measurements of \vpk\ match well with those in previous work,
including, for example, SN~2004et
\citep{Jerkstrand12}. \citet{Anderson14} found that \vpk\ of \hal\ for
a sample of SNe~IIP was typically in the range $-1000$ to +500~\kms, with
most of the features blueshifted and approaching zero velocity at
later epochs. This is consistent with what is found herein; compare
our Figure~\ref{f:vpk_hal} to their Figure~4. Opposite to \ltot\ and
\lpk\ discussed above, \vpk\ of \hal\ is found to be anticorrelated
with late-time $V$-band slope. That is, objects with slower/shallower
slopes tend to have larger values of \hal\ \vpk. \citet{Anderson14},
using measurements from earlier in the life of SNe~IIP (specifically, the
$V$-band decline rate during the plateau phase and the \vpk\ of \hal\ 
measured at $t = 30$~d past explosion), find a similar
anticorrelation. Note that we find no significant correlation or
anticorrelation between the \hal\ \vpk\ value and \mpk, \mplat, or the
late-time decline rates in the $B$, $R$, $I$, or pseudobolometric
light curves.

\subsection{Half Width at Half-Maximum Intensity}\label{ss:hwhm}

Our measurements indicate that the HWHM of the spectral features
investigated generally decrease with time, with a rapid decline at ages
earlier than \about300~d past explosion and a shallower decline
thereafter. This is similar to the results of \citet{Maguire12}, who
found relatively flat temporal evolution (for $300 < t < 600$~d past
explosion) of the HWHM of \hal, [\ion{O}{I}] $\lambda$6300,
[\ion{Ca}{II}] $\lambda$7291, and [\ion{Fe}{II}] $\lambda$7155. The
actual range of HWHM values that we measure for these features is also
mostly consistent with that of \citet{Maguire12}. 

For a few objects, including SN~2004dj with spectra having $t > 600$~d
past explosion, the HWHM increases at later times. This is primarily 
caused by the spectral features getting weaker and broader with time, which
leads to smaller \lpk\ values, but larger HWHM values. This evolution
is consistent with the findings of \citet{Milisavljevic12}, even
though most of their data are from much later epochs. Their work finds
a similar amount of decrease in \lpk\ of \hal\ to what is found in the
current study during the first 1--2~yr after
explosion. Furthermore, spectra from \citet{Milisavljevic12}, as well
as \citet{Blair15}, show \hal\ and [\ion{O}{I}] $\lambda\lambda$6300,
6364 profiles that are similar in appearance to those seen in our
oldest spectra. These works attribute this evolution of the spectral
profiles of SNe~IIP at late times to the SNe beginning their
transiation to the remnant phase. 

\begin{figure}
\centering
\includegraphics[width=3.5in]{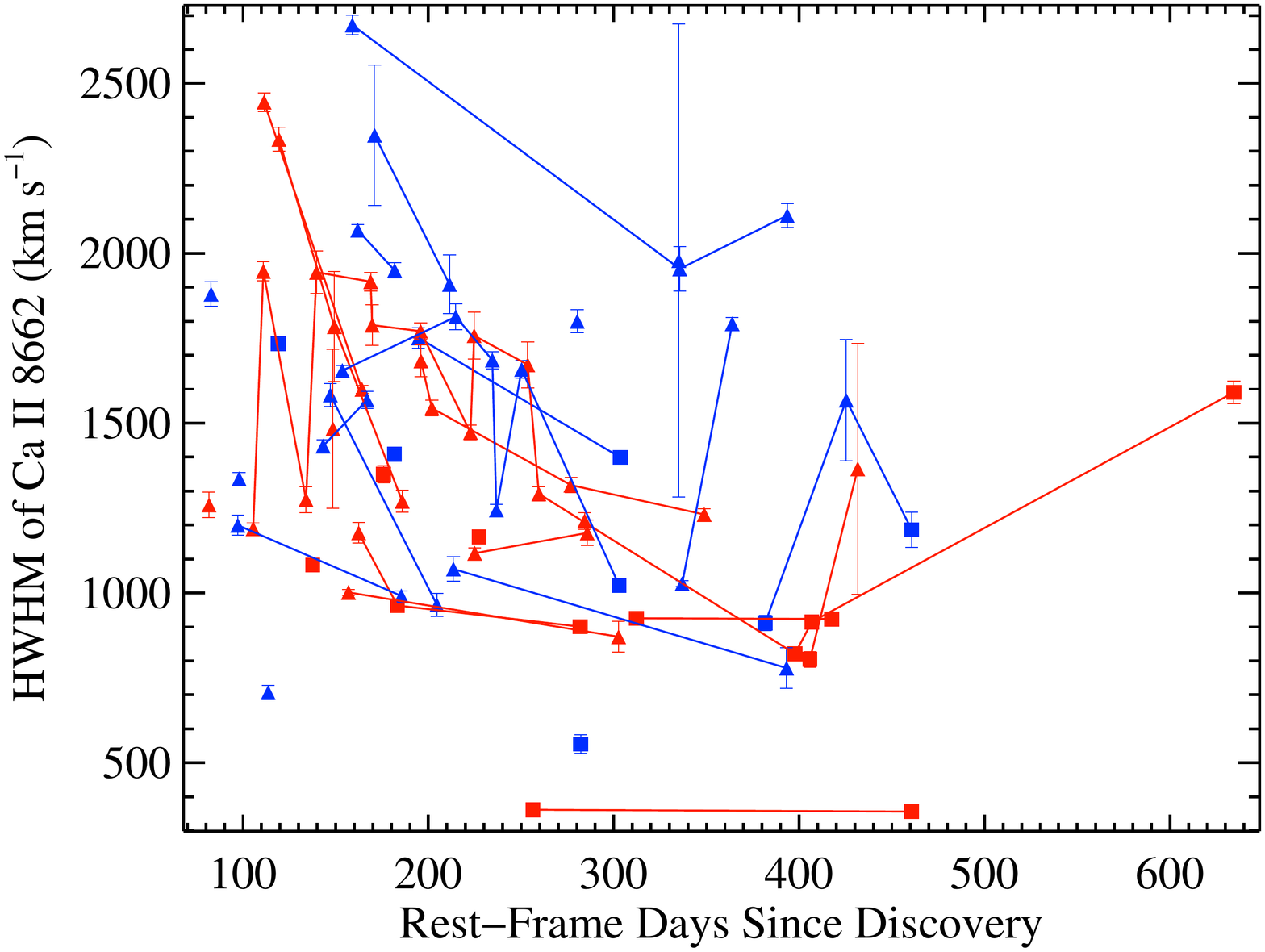} 
\caption{The HWHM of \ion{Ca}{II} $\lambda$8662 versus time. Blue
  points are SNe~IIP with \mplat\ brighter than or equal to
  $-16.3$~mag; red points have \mplat\ fainter than
  $-16.3$~mag. Squares represent profiles that are fit well by a
  single-Gaussian function; triangles represent profiles
  whose shapes are more complex (see Section~\ref{ss:shape} for more
  information). Spectra 
  of the same object are connected with solid
  lines.}\label{f:hwhm_ca8662_t_mplat}
\end{figure}

\begin{figure}
\centering
\includegraphics[width=3.5in]{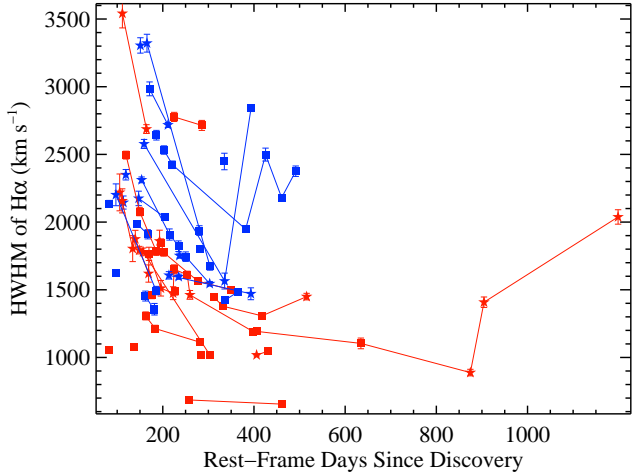} 
\caption{The HWHM of \hal\ versus time. Blue points are SNe~IIP with
  \mplat\ brighter than or equal to $-16.3$~mag; red points have
  \mplat\ fainter than $-16.3$~mag. Squares represent profiles that
  are fit well by a single-Gaussian function; stars represent profiles
  whose shapes are more complex (see Section~\ref{ss:shape} for
  more information). Spectra of the same object are connected with
  solid lines.}\label{f:hwhm_hal}
\end{figure}

The typical HWHM we find for the elements that originate from the
helium core of the progenitor star \citep[oxygen, calcium, and helium;
e.g.,][]{Dessart11} are \about1000--1200~\kms, except for 
\ion{Ca}{II} $\lambda$8662 which is closer to \about1500~\kms\ (see
Figure~\ref{f:hwhm_ca8662_t_mplat}). These 
values are all smaller than what was predicted by \citet{Dessart11},
but they do note that \ion{Ca}{II} $\lambda$8662 should have
larger HWHM values than the rest of the oxygen, calcium, and helium
lines since it is
formed from both the helium core and hydrogen envelope. The lowest
minimum HWHM values are measured for [\ion{Ca}{II}] and \ion{Ca}{II} 
(\about500--800~\kms), which implies that they are mixed down to the
lowest velocities and innermost radii of the ejecta. The next-lowest
minimum HWHM values are found in [\ion{O}{I}], \ion{O}{I}, and
\ion{He}{I}  (900--1000~\kms), followed by the hydrogen Balmer lines
which have the largest HWHM \citep[with most $>1300$~\kms; see
Figure~\ref{f:hwhm_hal} and compare to Figure~5
of][]{Maguire12}. 

Our measurements indicate that the HWHM of \hal\ and all oxygen
spectral features (but not other lines in the Balmer series or calcium
features) are larger in SNe~IIP with brighter \mpk\ and \mplat\ (see
Figure~\ref{f:hwhm_hal}). A KS test indicates the HWHM values of these
features for objects with \mpk\ brighter than $-16.6$~mag
(or \mplat\ brighter than $-16.3$~mag) are statistically different 
than fainter objects ($p =
0.01$--0.03). \citet{Spiro14} came to a similar conclusion at slightly
earlier epochs, as they found broader spectral feature profiles at the
end of the plateau phase in objects with brighter plateaus. 

As mentioned above, the median HWHM values are also found to be larger
for objects with larger/steeper $V$-band slopes, especially in the
strongest emission lines (i.e., \hal, [\ion{O}{I}]
$\lambda\lambda$6300, 6364, and [\ion{Ca}{II}]
$\lambda\lambda$7291, 7324). Thus, broad emission lines of [\ion{O}{I}]
and \hal\ indicate luminous light-curve plateaus and steep $V$-band
decline rates, which is seen in the models of
\citet{Dessart13}. Furthermore, theoretical models indicate that for a
given explosion energy large HWHM values of [\ion{O}{I}] features
come from a large progenitor mass and radius
\citep{Dessart10,Dessart11}, while the HWHM of \hal\ increases with
greater mixing within the SN ejecta \citep{Dessart13}. Therefore,
SNe~IIP with broader [\ion{O}{I}] and \hal\ emission lines are also
likely to have larger progenitors and ejecta with more
thoroughly mixed hydrogen and oxygen layers.

\subsection{Flux Ratios}\label{ss:ratios}

In addition to individual spectral features discussed previously, we
follow what has been done in other late-time SN~IIP studies and also
investigate some flux ratios of pairs of emission lines. We calculated
the Balmer decrement (i.e., the ratio of \hal\ to H$\beta$, which were
all found to be $>3$), as well as the ratio between [\ion{O}{I}]
$\lambda\lambda$6300, 6364 and H$\beta$, but no significant
correlations were found with any other observables. 

\begin{figure*}
\centering$
\begin{array}{cc}
\includegraphics[width=3.4in]{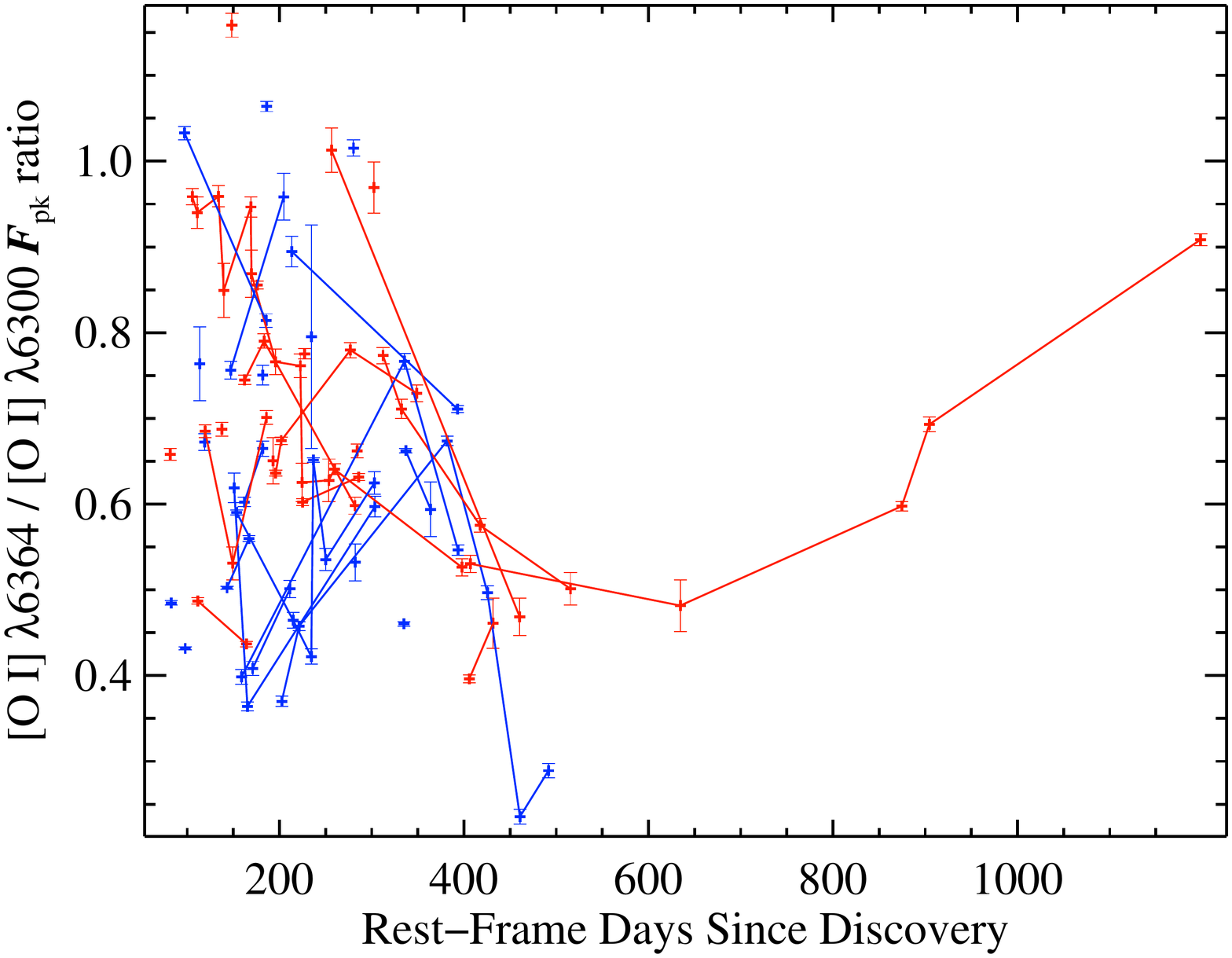} &
\includegraphics[width=3.4in]{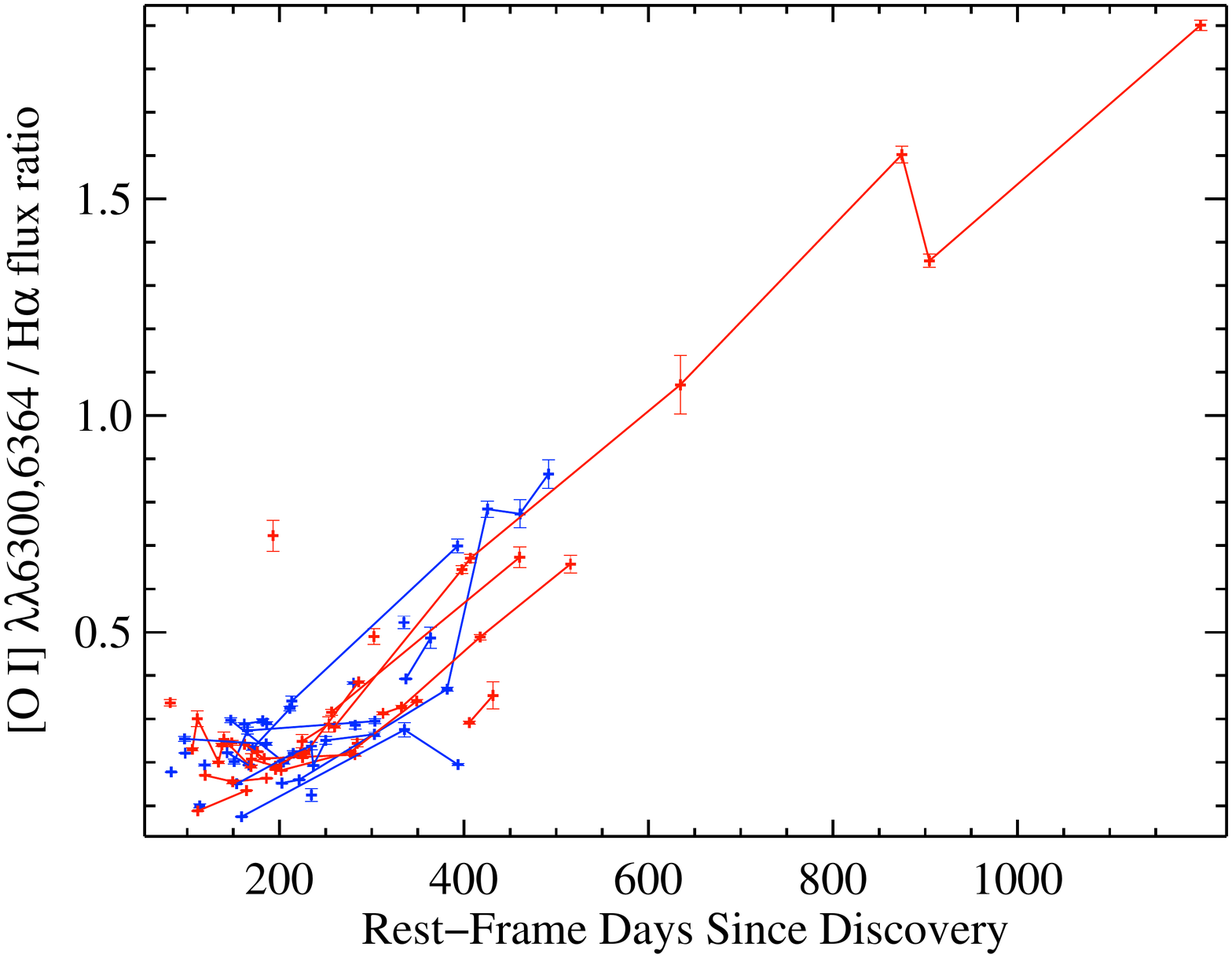} \\
\includegraphics[width=3.4in]{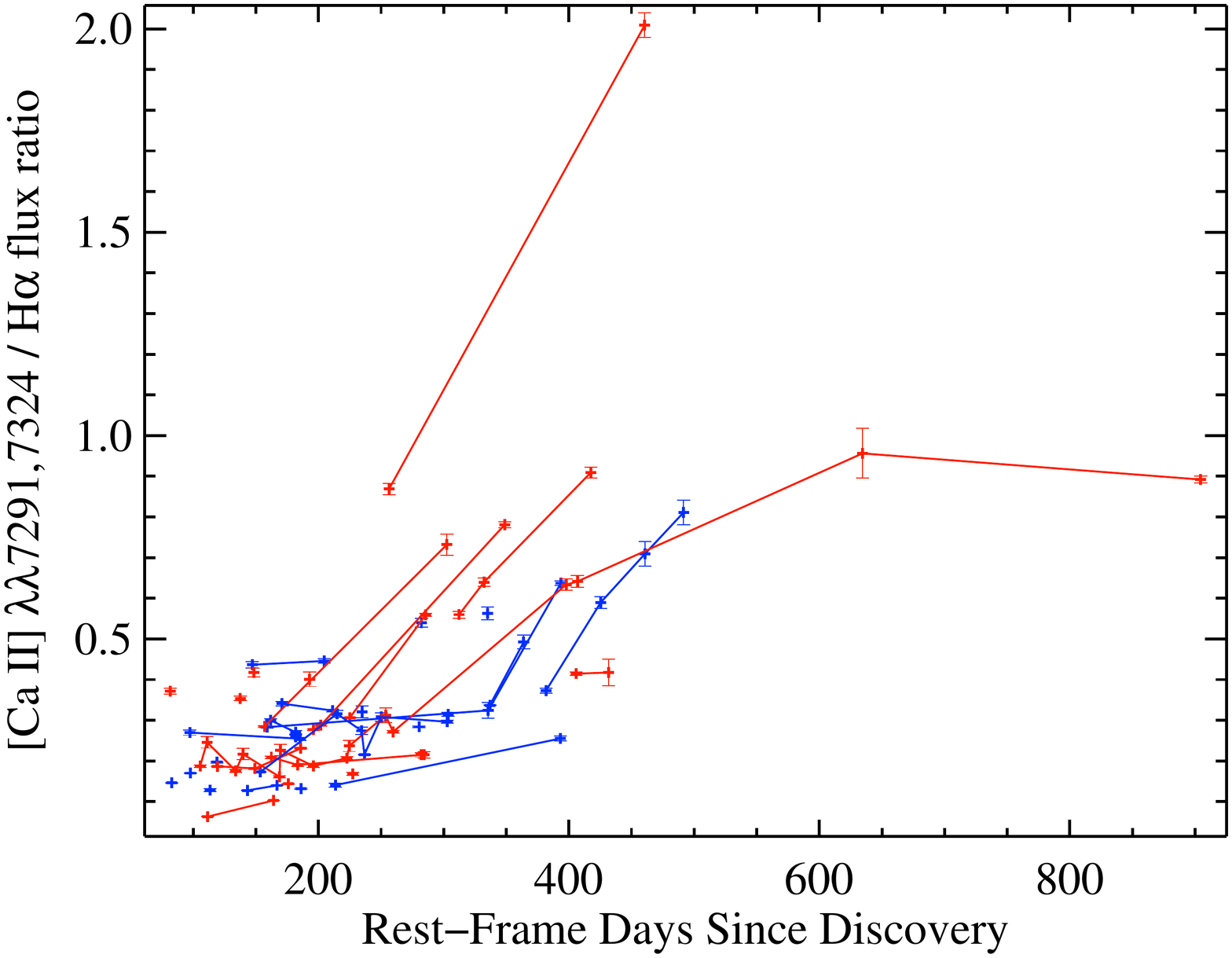} &
\includegraphics[width=3.4in]{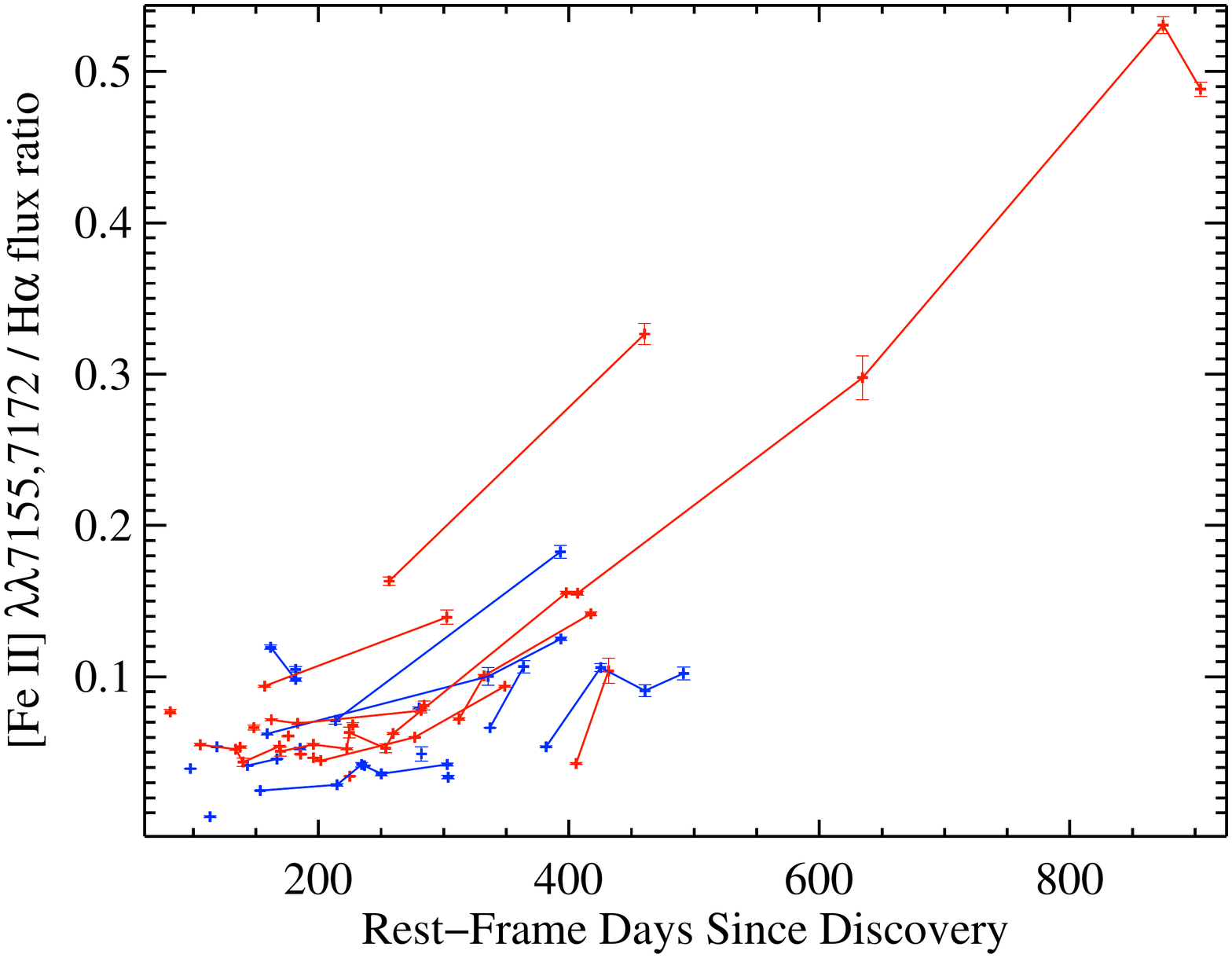} \\
\end{array}$
\caption{Various flux ratios versus time: the [\ion{O}{I}] peak flux
  ratio (top left), the [\ion{O}{I}] $\lambda\lambda$6300, 6364 doublet
  to \hal\ ratio (top right), the [\ion{Ca}{II}]
  $\lambda\lambda$7291, 7324 doublet to \hal\ ratio (bottom left), and
  the [\ion{Fe}{II}] $\lambda\lambda$7155, 7172 doublet to \hal\ ratio
  (bottom right). Blue points are SNe~IIP with \mplat\ brighter than
  or equal to $-16.3$~mag; red points have \mplat\ fainter than 
  $-16.3$~mag. Spectra of the same object are connected with solid
  lines.}\label{f:ratios}
\end{figure*}

The red-to-blue peak flux ratio of [\ion{O}{I}]
$\lambda\lambda$6300, 6364, defined as \fpk\ of [\ion{O}{I}]
$\lambda$6364 divided by \fpk\ of [\ion{O}{I}] $\lambda$6300
\citep[e.g.,][]{Chugai92,Maguire12}, is calculated for 89 of our 91
spectra. As seen in the top left panel of Figure~\ref{f:ratios}, this
ratio mostly decreases from \about1.0 (at $t \approx 100$~d) to
\about0.4--0.5 (at $200 \la t \la 500$~d). There is possibly an
increase at even later epochs, but there is only one object
(SN~2004dj) in our sample at such late times. In the optically thick
regime this ratio should approach 1, while in the optically thin
regime it should approach 1/3. Our measured values are mostly within
these limits, and deviations $>1$ could be caused by the blending of
nearby spectral features or electron scattering and clumpiness
\citep{Chugai92}. Figure~12 of \citet{Maguire12} displays less scatter
than our larger sample, but we both find a relatively smooth decrease
in the ratio with time, especially for $t > 200$~d. The [\ion{O}{I}]
peak flux ratio was also calculated from observations and modeled by
\citet{Spyromilio91}. The ratios they measure are similar to the
values we find, and nearly all of our measurements appear to be fit by
their models if one sets the velocity extent of the O-emitting region
to 2000--3000~\kms, the temperature to \about2000~K, the filling
factor to 0.01--0.1, and the mass of \ion{O}{I} to \about1~\msun.

The ratio of \ftot\ of the [\ion{O}{I}] $\lambda\lambda$6300, 6364
doublet to \ftot\ of \hal\ was calculated and is shown in the 
top-right panel of 
Figure~\ref{f:ratios}. The ratio is relatively constant in time for $t
< 200$~d and then increases (mostly) monotonically thereafter. 
In the bottom-left and bottom-right panels of Figure~\ref{f:ratios} we
present the ratio of the [\ion{Ca}{II}] $\lambda\lambda$7291, 7324
doublet to \hal\ and the ratio of
[\ion{Fe}{II}] $\lambda\lambda$7155, 7172 to \hal, respectively. Much
like the [\ion{O}{I}] $\lambda\lambda$6300, 6364 to \hal\ ratio, these
both show relatively constant values with time for $t < 200$--250~d
and then an increase at later epochs. 
The results for all of these ratios are very similar to those of
\citet{Maguire12}, although they do not present data as early as in
our sample and thus do not observe the epoch of nearly constant ratios
at $t < 200$--250~d. 

For completeness, and to compare with \citet{Maguire12}, we also
computed the [\ion{Ca}{II}] $\lambda\lambda$7291, 7324 to [\ion{O}{I}]
$\lambda\lambda$6300, 6364 ratio, the
[\ion{Fe}{II}] $\lambda\lambda$7155, 7172 to [\ion{O}{I}]
$\lambda\lambda$6300, 6364 ratio, and the
[\ion{Fe}{II}] $\lambda\lambda$7155, 7172 to [\ion{Ca}{II}]
$\lambda\lambda$7291, 7324 ratio. All of these ratios showed large
scatter versus time. Furthermore, they were mostly consistent with the
range of measured values and general temporal behaviour seen by 
\citet{Maguire12}.

\subsection{Spectral Feature Shapes}\label{ss:shape}

As mentioned in Section~\ref{s:measuring}, and denoted by the shapes of
the data points in some of the previous figures, each spectral feature
in our sample was assigned a descriptor of its overall visual
shape or appearance. As most of our spectra have resolution better than
14~\AA\ (\about650~\kms), the observed shapes should reflect the
intrinsic shapes of 
moderately strong emission features \citep{Maguire12}. Consistent with
the work of \citet{Maguire12}, we found examples of single-peaked
profiles and multi-peaked profiles, but no evidence was found for
flat-topped profiles, which would indicate ejecta layers that are not
well-mixed. Of the profiles that were marked as multi-peaked, most
appeared to be double-peaked and ones that had more than two distinct
peaks were found to be the result of other emission features blended
with the lines of interest. 

Of the double-peaked profiles, some were caused by noise in the data,
some by nearby blended emission lines (as in the
multi-peaked profiles), and some by narrow emission from
the host galaxy of the SN~IIP. No narrow emission lines from the SNe
themselves were detected in any of our spectra. This type of emission
profile could arise from late-time interaction with circumstellar
media in the same manner as so-called Type~IIn-P SNe
\citep[e.g.,][]{Mauerhan13}. Removing these noisy, blended, or
contaminated spectra left only profiles that are double-peaked owing 
to two separate emission peaks of the same spectral feature in the SN
ejecta. 

\begin{figure}
\centering
\includegraphics[width=3.6in]{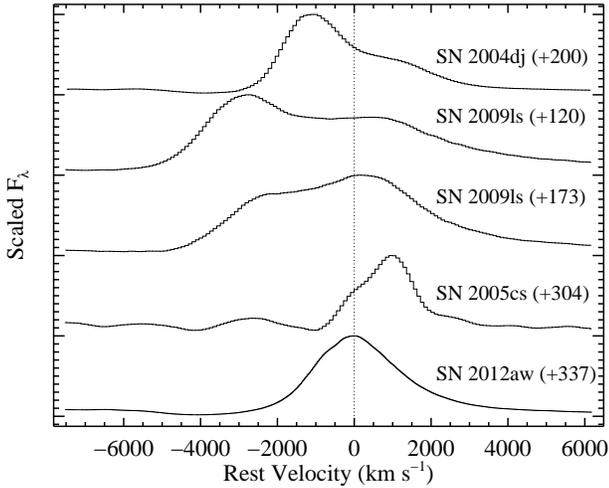} 
\caption{Various \hal\ profiles. Each spectrum is labeled with its
  object name and rest-frame age relative to explosion and has been 
  corrected for its host-galaxy recession velocity and Galactic
  reddening using the values listed in Table~\ref{t:objects}. The
  dotted vertical line is the zero velocity of \hal. The top two
  spectra show a blueshifted peak with a red shoulder, the next two
  spectra exhibit a redshifted peak with a blue shoulder, and the bottom
  spectrum has a single-peaked, mostly symmetric profile that peaks
  at the rest wavelength of \hal.}\label{f:hal_profiles}
\end{figure}

Since \hal\ is the strongest emission line in the spectra of SNe~IIP, 
it has the highest S/N and thus two distinct peaks can be seen most 
easily in this feature. We find 8 SNe~IIP with \hal\ profiles that
have blueshifted peaks with a red shoulder (i.e., a second,
weaker, redshifted peak). One of these objects, SN~2009ls evolves to
the opposite profile (i.e., a redshifted peak with a blue
shoulder) between our two spectra of that object (120~d past explosion
to 173~d past explosion). Three other SNe~IIP in our sample exhibit a
redshifted peak with a blue shoulder. All of the spectra that  
show these double-peaked \hal\ profiles are younger than \about300~d
past explosion. Both \hal\ profiles of SN~2009ls, as well as two other
double-peaked \hal\ profiles and a single-peaked \hal\ profile (for
comparison), are plotted in Figure~\ref{f:hal_profiles}.

Asymmetric, multi-peaked, or otherwise complex profiles were searched
for in other emission lines, but they are harder to distinguish 
than in \hal\ owing to either their relatively lower flux or
blending with other nearby features, or both. The [\ion{O}{I}]
$\lambda\lambda$6300, 6364, [\ion{Ca}{II}] $\lambda\lambda$7291, 7324,
and [\ion{Fe}{II}] $\lambda\lambda$7155, 7172 doublets are all
relatively strong spectral features discussed above, but since they
are all doublets it is difficult to identify multiple, distinct
emission peaks. No convincing multi-peaked profiles were observed in
the [\ion{Ca}{II}] $\lambda\lambda$7291, 7324 or [\ion{Fe}{II}]
$\lambda\lambda$7155, 7172 features. On the other hand, 7 SNe~IIP
showed tentative evidence of blueshifted peaks and red shoulders
in their [\ion{O}{I}] $\lambda\lambda$6300, 6364 profiles.

Of the objects with asymmetric \hal\ profiles, two (SNe~1988H and
2004dj) showed the same profile shape in [\ion{O}{I}]
$\lambda\lambda$6300, 6364. Furthermore, SNe~IIP with blueshifted
\hal\ peaks tended to have the most negative peak velocities of
[\ion{O}{I}] $\lambda\lambda$6300, 6364 and [\ion{Fe}{II}]
$\lambda\lambda$7155, 7172. This is indicative of a blueshifted peak
in these features, possibly with a redshifted shoulder, but the
asymmetric profile is too weak or too blended to be visually
confirmed. Other than this, all objects with asymmetric profiles
appear to have typical spectral and photometric observables.
This result --- that emission lines of different ions in the same spectrum
tend to have the same overall profile shape --- is consistent with
previous work \citep{Maguire12}.

At very late times (i.e., $t \ga 400$~d), asymmetric or double-peaked
profiles in SN~IIP spectra are sometimes attributed to the presence of
dust \citep[e.g.,][]{Jerkstrand15}, but all of the spectra in the current
sample that show these sorts of profiles are younger than this. At the
epochs in question ($100 \la t \la 300$~d), \citet{Jerkstrand15} state
that dust will only have a ``small effect'' on the optical/NIR spectra
of SNe~IIP. Instead, it has been proposed that asymmetric $^{56}$Ni
ejection, possibly bipolar in shape, is responsible for the
asymmetric profiles seen at these epochs \citep[e.g.,][]{Chugai05}. In
fact, the strange case of the \hal\ profiles of SN~2009ls mentioned
above could be explained by a bipolar $^{56}$Ni distribution with a
time-variable covering fraction. In this situation, the observed area
covered by the $^{56}$Ni changes with time such that the approaching
lobe of $^{56}$Ni is observed first (giving rise to the blueshifted
\hal\ peak seen in the first spectrum), then at later times emission
from that lobe weakens as the receding lobe of $^{56}$Ni becomes
visible (leading to the redshifted \hal\ peak in the second
spectrum). It is not obvious, however, how the idea of bipolar
$^{56}$Ni ejection can explain the prevalence, by a factor of \about2,
of blueshifted peaks over redshifted peaks seen in our data. 

For most of the objects where we detect asymmetric \hal\ profiles,
previous work has not specifically commented on the profile
shape. The present study, however, is mostly consistent with examples 
in the literature that have investigated profile shapes in late-time
SNe~IIP spectra. SN~1988H exhibited many asymmetric profiles at
\about400~d past explosion \citep{Turatto93}, which matches our
detection of such profiles in multiple emission lines at $t = 140$~d
past explosion. We find that the well-studied SN~2004dj has a
blueshifted \hal\ peak with a red shoulder in spectra at $136 < t <
438$~d after explosion, which was also observed by \citet{Chugai05}
and \citet{Meikle11}. Furthermore, asymmetry in the ejecta has also
been observed via spectropolarimetry of SN~2004dj \citep{Leonard06},
consistent with bipolar $^{56}$Ni ejection. SN~2005cs is one of the
few objects that shows the opposite \hal\ profile (i.e., a redshifted
peak with a blue  shoulder), and it is seen in both spectra of this
object in our sample ($t = 158$ and 304~d past
explosion). \citet{Pastorello09} detect the same \hal\ profile shape
in spectra obtained at similar epochs.

Blueshifted peaks with red shoulders are possibly seen in the
\hal\ profiles of spectra of SN~1999em at $200 < t < 300$~d after
explosion \citep{Elmhamdi03b}, but we find no compelling evidence of
an asymmetric \hal\ profile in our spectra from slightly later epochs
($t = 317$ and 337~d past explosion). At $t > 300$~d past explosion,
dust was likely present in SN~2004et \citep{Kotak09}, and there are
indications of a blueshifted \hal\ peak \citep{Sahu06}; however, we do
not detect an asymmetric \hal\ profile in our spectra of this object 
at $202 < t < 355$~d past explosion.


\section{Comparisons to Theoretical Models}\label{s:models}

In the following subsections, we compare our late-time spectral data
of SNe~IIP to two recent studies that presented sets of theoretical
spectra: \citet{Dessart13} and \citet{Jerkstrand14}. In general, the
vast majority of our spectra match only moderately well to models from
the former, but match quite well to models from the latter.

\subsection{\citet{Dessart13} Models}\label{ss:dessart13}

\citet{Dessart13} model 15~\msun\ stars as the progenitors of
SNe~IIP. Adjustable parameters in their models include the mixing
length, the amount of convective overshoot, the amount of stellar
rotation, and the progenitor metallicity. They produce and then evolve
many pre-SN progenitors and then model the SN~IIP ejecta from early to
late times. As pointed out by \citet{Dessart13}, one of the
shortcomings of their model spectra is that \ion{He}{I} lines are
overproduced (especially the 7065~\AA\ line) relative to the
observations to which they compare. In the present work we clearly
detect \ion{He}{I} $\lambda$7065 emission in about half of our 
late-time spectra, although it is usually weaker than what is
predicted by the models of \citet{Dessart13}.

We compared every spectrum in our sample to six models run by
\citet{Dessart13} which varied progenitor metallicity ($Z$) and mixing
length parameter ($\alpha$). Late-time spectra at a variety of epochs
were produced for each model, so the spectra in our sample were
compared to each model spectrum at the closest epoch. Using visual
inspection and a basic cross-correlation algorithm, the model that was
most consistent with each spectrum was chosen. Then the model that
best fit the majority of the spectra of a given object was deemed the
model most consistent with that SN~IIP.  

Usually, no model fit an individual spectrum very well. Aside from the
models showing too much helium emission, as mentioned above, they also
sometimes incorrectly predict the relative peak fluxes of the
strongest lines (i.e., hydrogen, oxygen, and calcium). Furthermore,
there was often more than one model that matched an individual object
relatively well. Specifically, models where the only difference was
metallicity and the values of $Z$ were in the middle of the range
tested (i.e., 0.008--0.020) looked very similar. Our final analysis
indicates that 5 objects do not match any model reasonably well,
while half of the SNe~IIP in our dataset are consistent with models
with either $Z = 0.002$ or $\alpha = 3$. In addition, it 
appears that most of the objects are consistent with models with
relatively low metallicity ($Z \le 0.01$). Of the objects in our sample
that have published metallicity measurements at the SN site, we find
the metallicities to be in the range $0.003 \la Z \la 0.014$, with a
typical value of \about0.011 \citep[e.g.,][]{Anderson16,Taddia16}, for
$Z_\odot = 0.0134$ \citep{Asplund09}. Thus, SNe~IIP do tend to be
found in low (i.e., subsolar) metallicity regions, but perhaps not
quite {\it as} low as the \citet{Dessart13} models would suggest.  

\begin{figure*}
\centering$
\begin{array}{cc}
\includegraphics[width=3.6in]{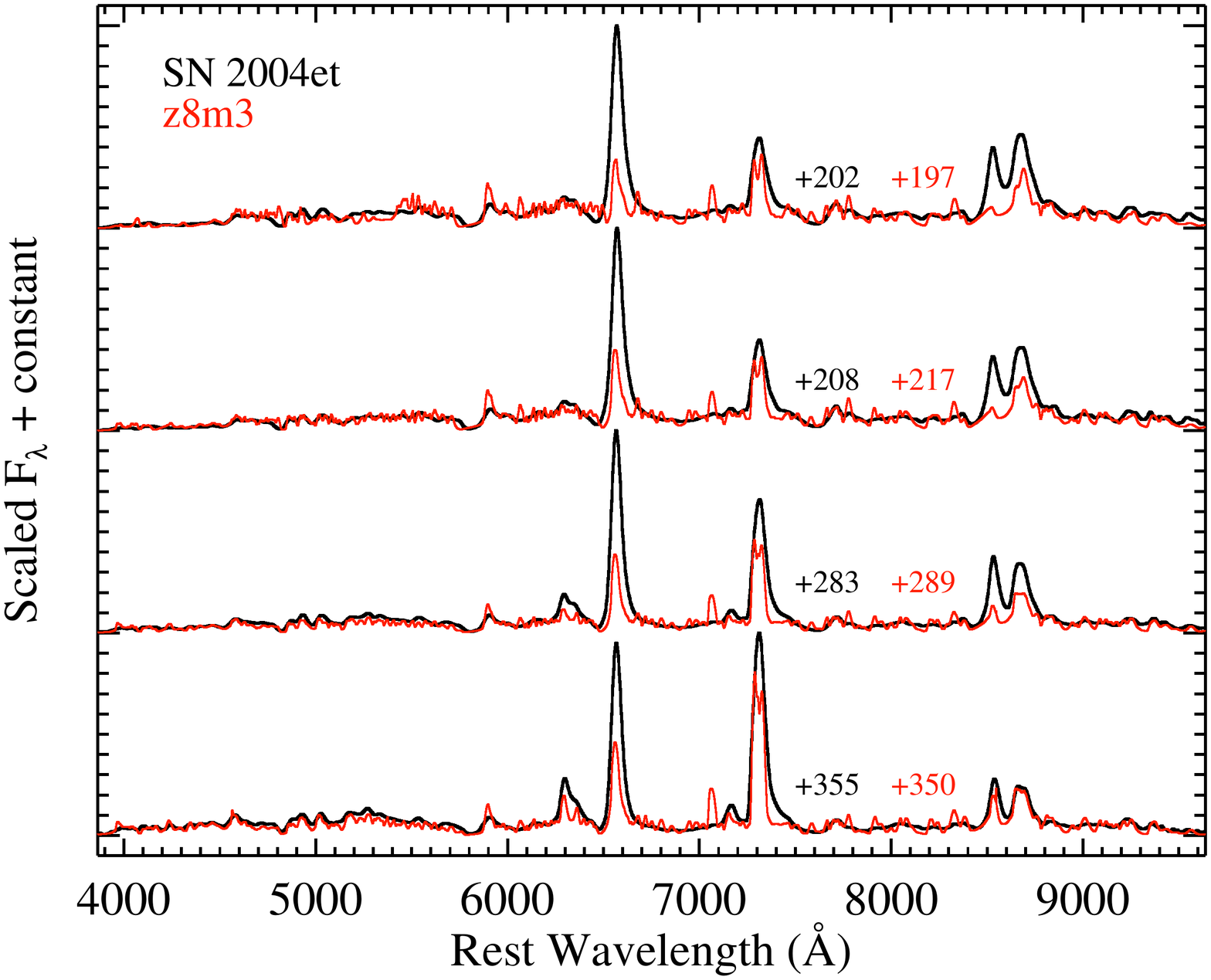} &
\includegraphics[width=3.6in]{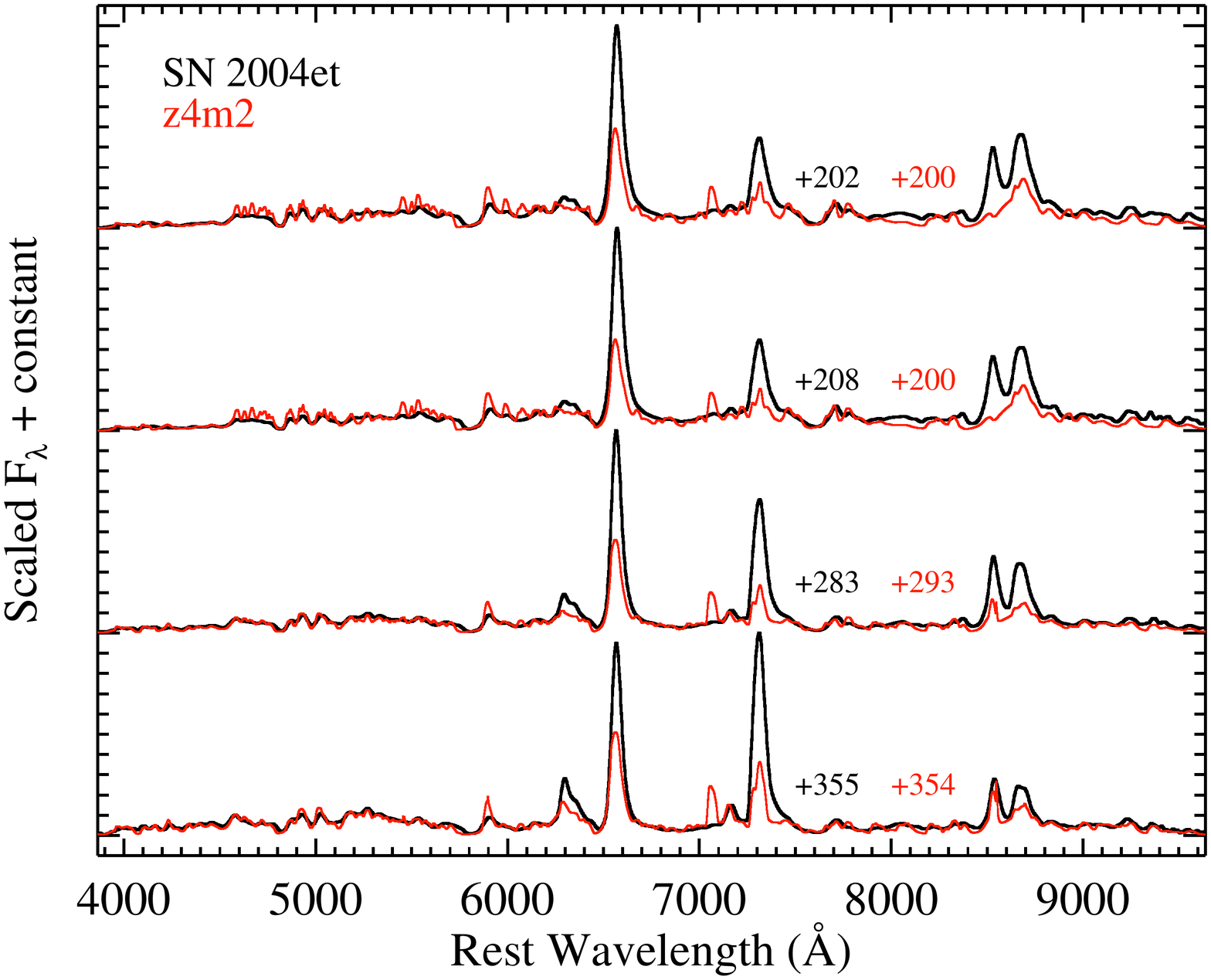} \\
\includegraphics[width=3.6in]{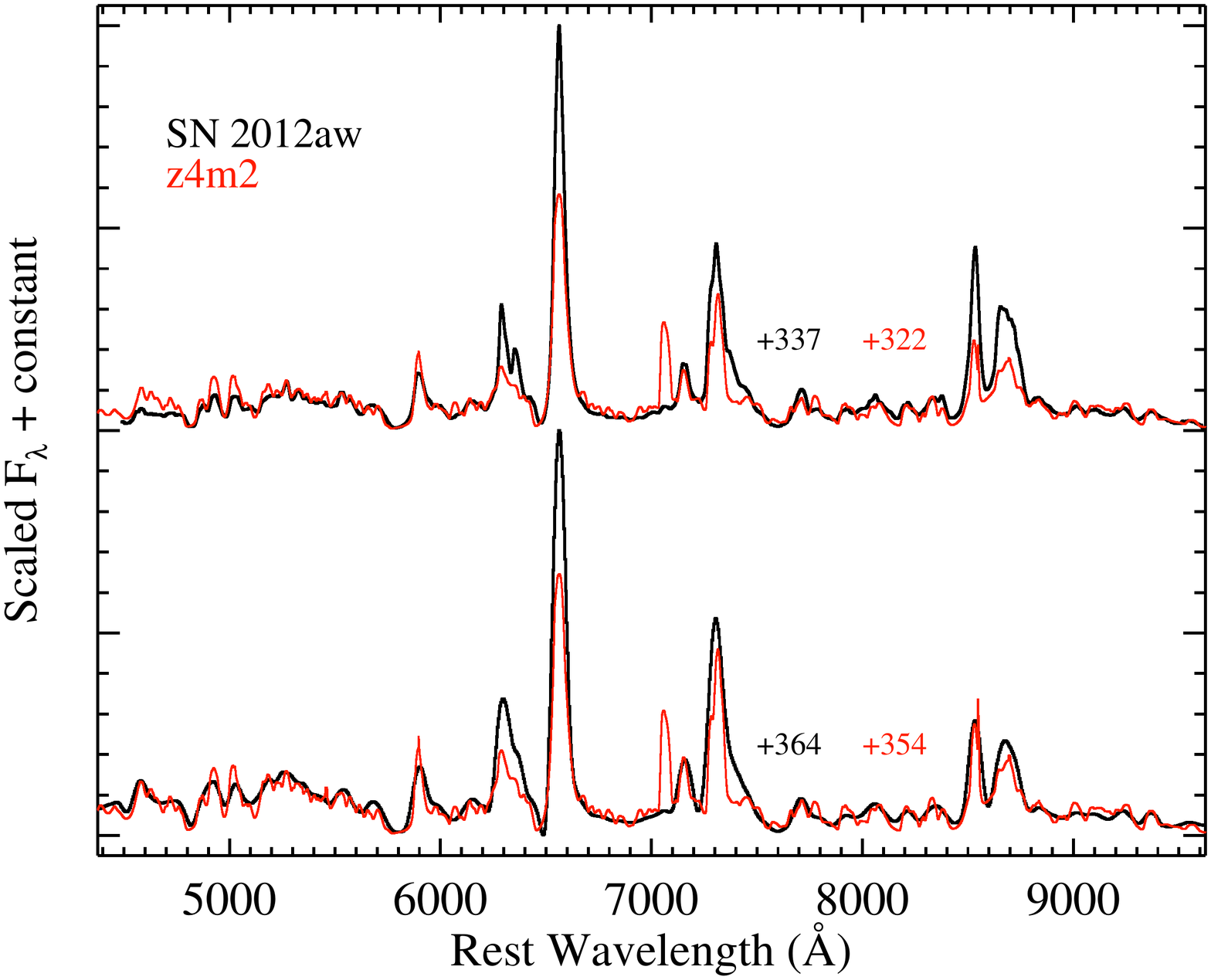} &
\includegraphics[width=3.6in]{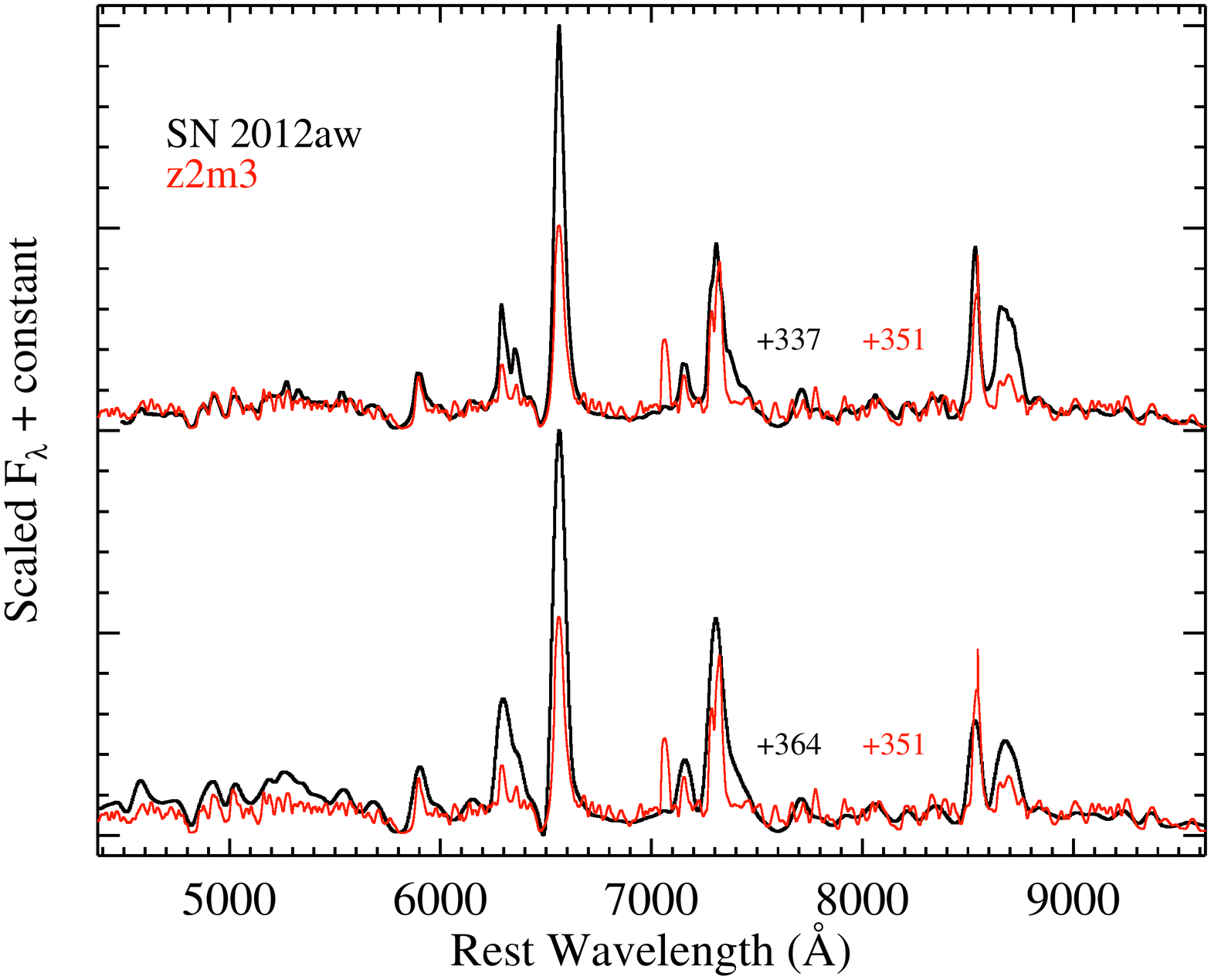} \\
\end{array}$
\caption{A comparison between two SNe~IIP and 
  models from \citet{Dessart13}. SN~2004et (top row) is consistent
  with the ``z8m3'' model ($Z = 0.008$ and $\alpha = 1.6$, top-left
  panel) and less so with the ``z4m2'' model ($Z = 0.040$ and $\alpha
  = 1.6$, top-right panel). On the other hand, SN~2012aw (bottom
  row) resembles the ``z4m2'' model ($Z = 0.040$ and $\alpha
  = 1.6$, bottom-left panel) and is less consistent with the ``z2m3''
  model ($Z = 0.002$ and $\alpha = 1.6$, bottom-right panel). Each
  spectrum is labeled with its rest-frame age relative 
  to explosion (black) and the age of the overplotted model spectrum
  relative to explosion (red). All data have been corrected for
  host-galaxy recession velocity and Galactic reddening using the 
  values listed in Table~\ref{t:objects}.}\label{f:dessart_models} 
\end{figure*}

Figure~\ref{f:dessart_models} shows SN~2004et (top row) and
SN~2012aw (bottom row) overplotted with models from
\citet{Dessart13}. SN~2004et is consistent with their ``z8m3''
model ($Z = 0.008$ and $\alpha = 1.6$; top-left panel), even though
\hal, oxygen, and calcium emission features are slightly weaker in the
model while the helium features are too strong. A less good match is
found with their ``z4m2'' model ($Z = 0.040$ and $\alpha = 1.6$) and
is shown in the top-right panel of the figure. SN~2012aw resembles
the ``z4m2''  model (in the bottom-left panel of
Figure~\ref{f:dessart_models}) of \citet{Dessart13}, which comes from
a star with $Z = 0.040$ and $\alpha = 1.6$, and is one of the few
objects that is consistent with a higher value of $Z$. Like the
comparison to SN~2004et, this model spectrum has \hal, oxygen, and
calcium features that are a little too weak and helium features that
are too strong. Their ``z2m3'' model ($Z = 0.002$ and $\alpha = 1.6$)
is less consistent with SN~2012aw and is shown in the bottom-right
panel.

\subsection{\citet{Jerkstrand14} Models}\label{ss:jerkstrand14}

\citet{Jerkstrand14} present late-time spectra of SN~2012aw, along
with theoretical spectra derived from stellar evolution and explosion
models. They produce late-time spectra of a variety
models with a nearly constant explosion energy that we can compare to
all of the objects in our sample. In general, the model spectra of
\citet{Jerkstrand14} show stronger [\ion{O}{I}]
$\lambda\lambda$6300, 6364 emission for larger progenitor masses while
all emission lines tend to weaken with time.

Using the same procedure outlined in Section~\ref{ss:dessart13} above,
we compared every spectrum in our sample to the model spectrum at the
closest epoch from each of the four progenitor masses (12, 15, 19,
and 25~\msun) modeled by \citet{Jerkstrand14}. The model that was most
consistent with each object was again determined by which model best fit
the majority of the spectra of each object. Even more so than with the models of
\citet{Dessart13}, there were many cases where two models
from \citet{Jerkstrand14} resembled the same spectrum or object
equally well. Thus, we caution against making precise interpretations of
progenitor mass from these model comparisons.

While the model spectra of \citet{Jerkstrand14} often match quite well
to our data, there are some issues. For example, as pointed out by
\citet{Jerkstrand14}, their models overproduce \hal\ emission and
underproduce the \ion{Ca}{II} NIR triplet, which we confirm in many of
the comparisons to our spectral sample. Furthermore, their models
sometimes overproduce helium emission, like the models of
\citet{Dessart13} discussed above, and occasionally incorrectly
predict the strength of the [\ion{Ca}{II}] $\lambda\lambda$7291, 7324
doublet.

\begin{figure*}
\centering$
\begin{array}{cc}
\includegraphics[width=3.6in]{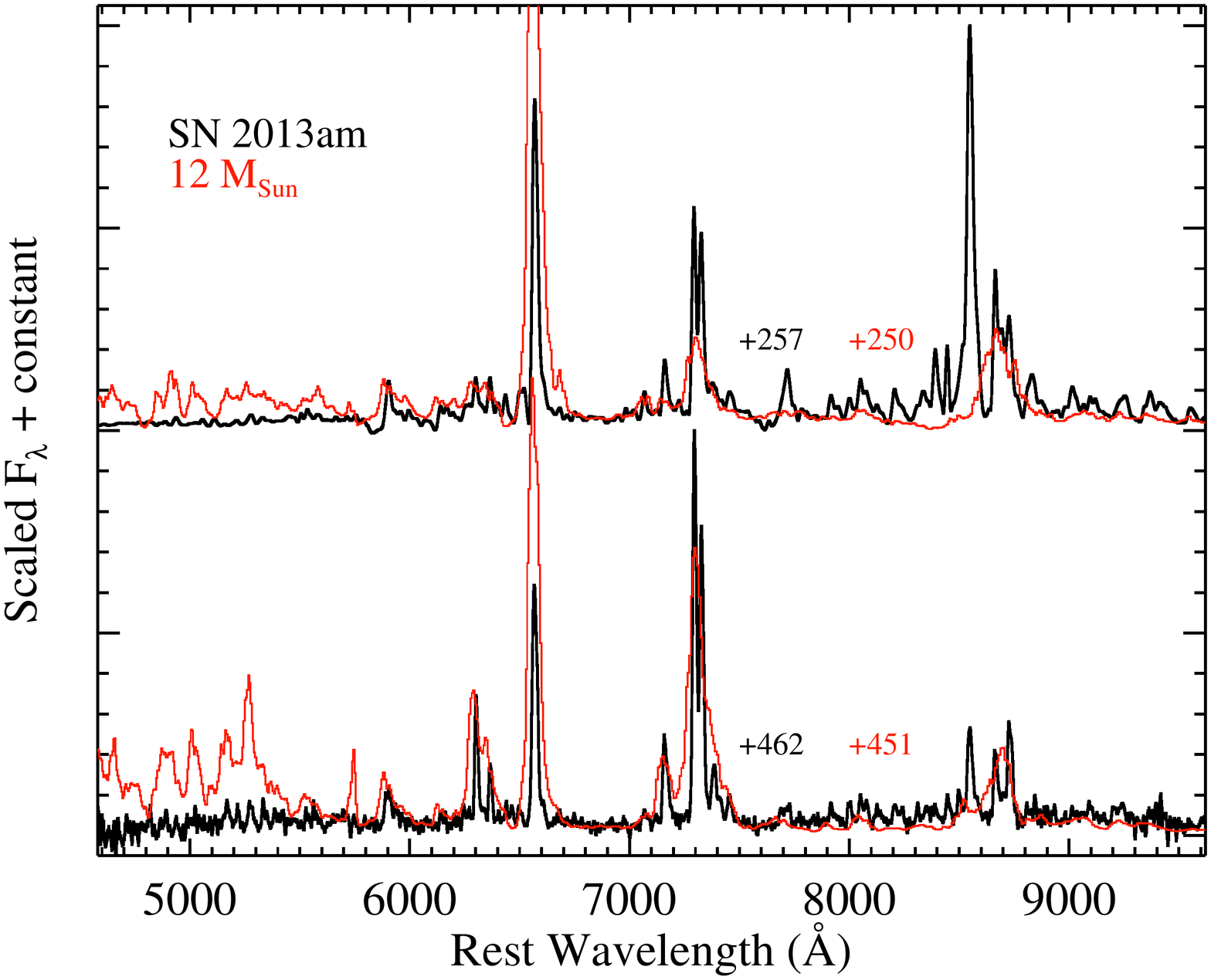} &
\includegraphics[width=3.6in]{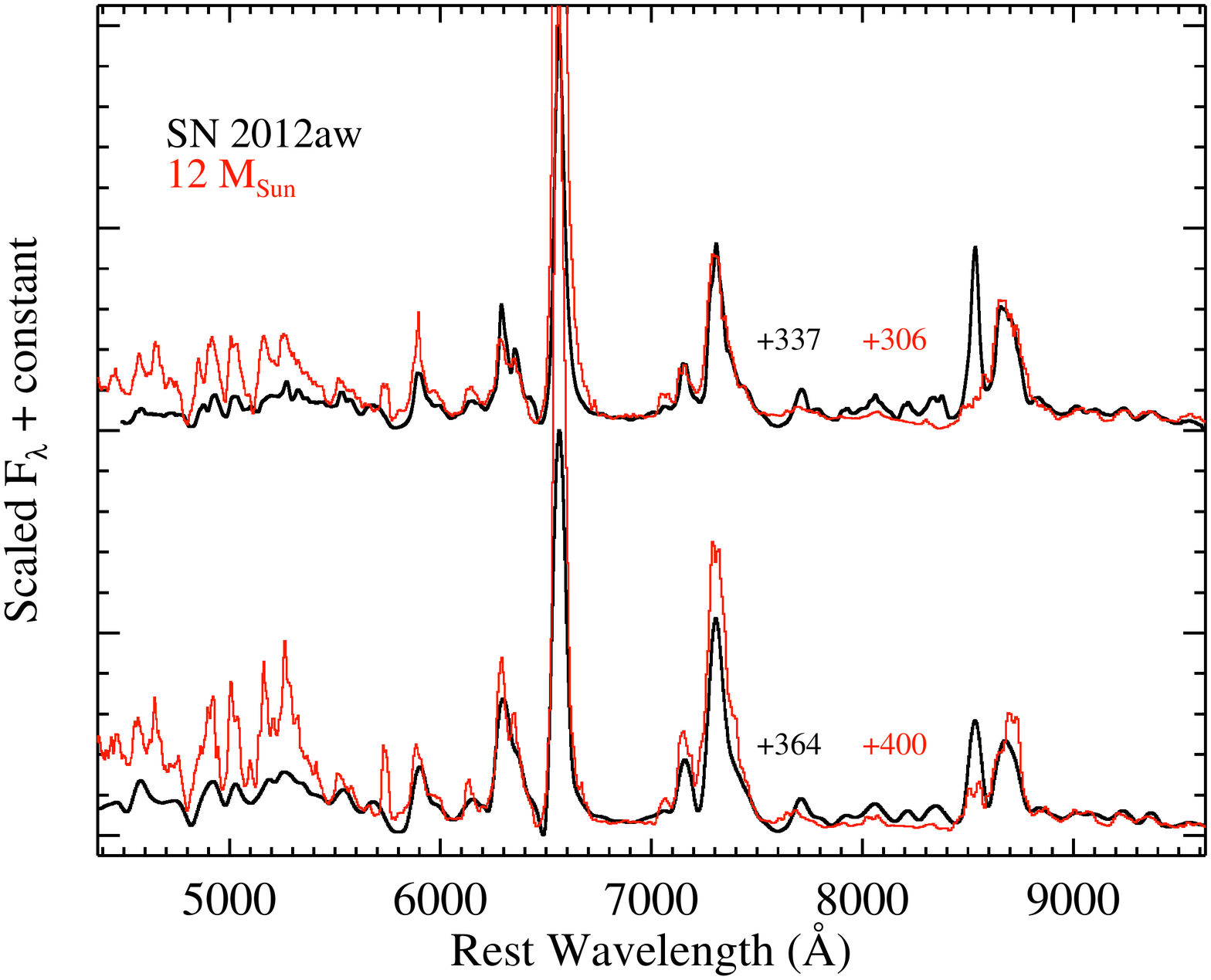} \\
\includegraphics[width=3.6in]{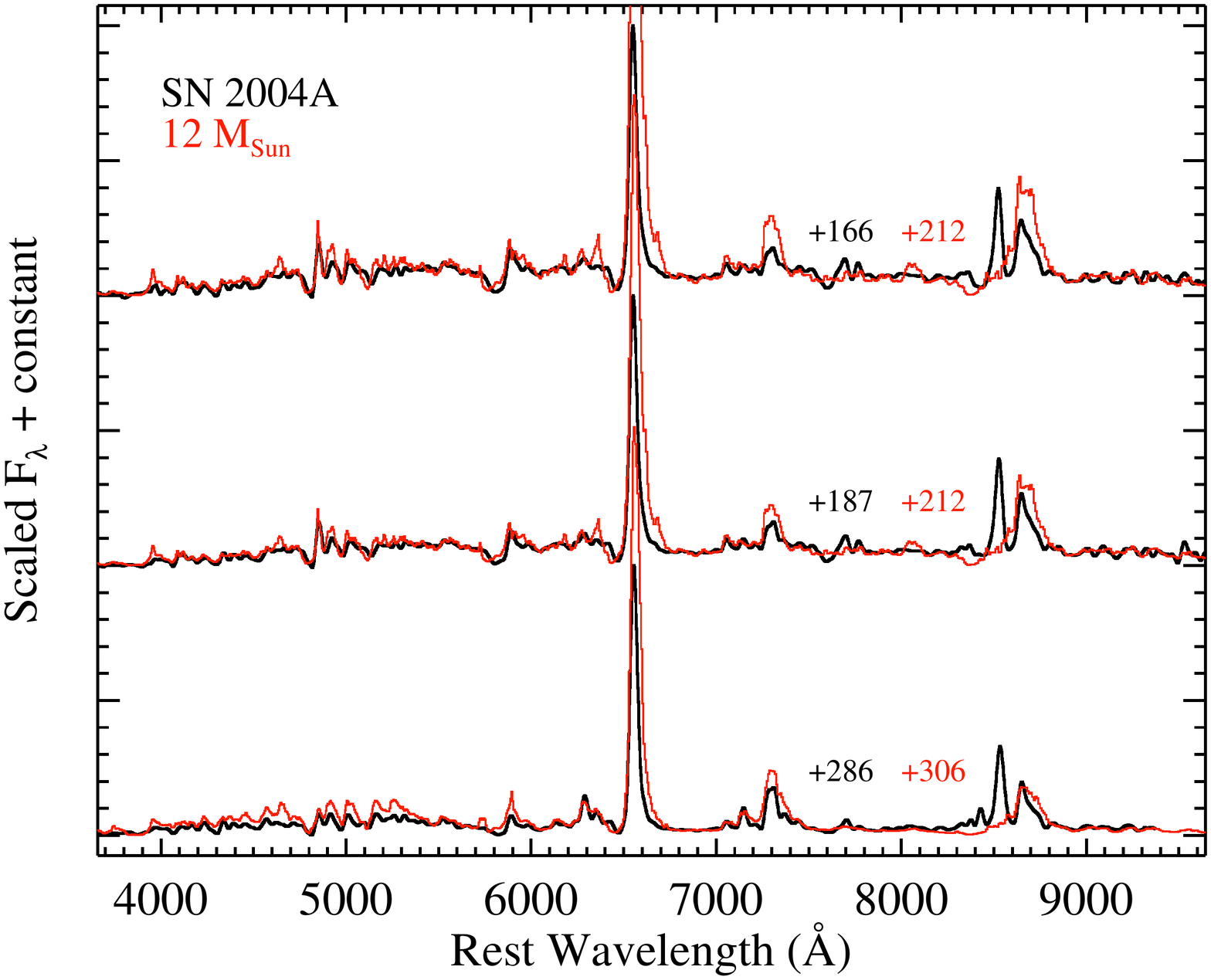} &
\includegraphics[width=3.6in]{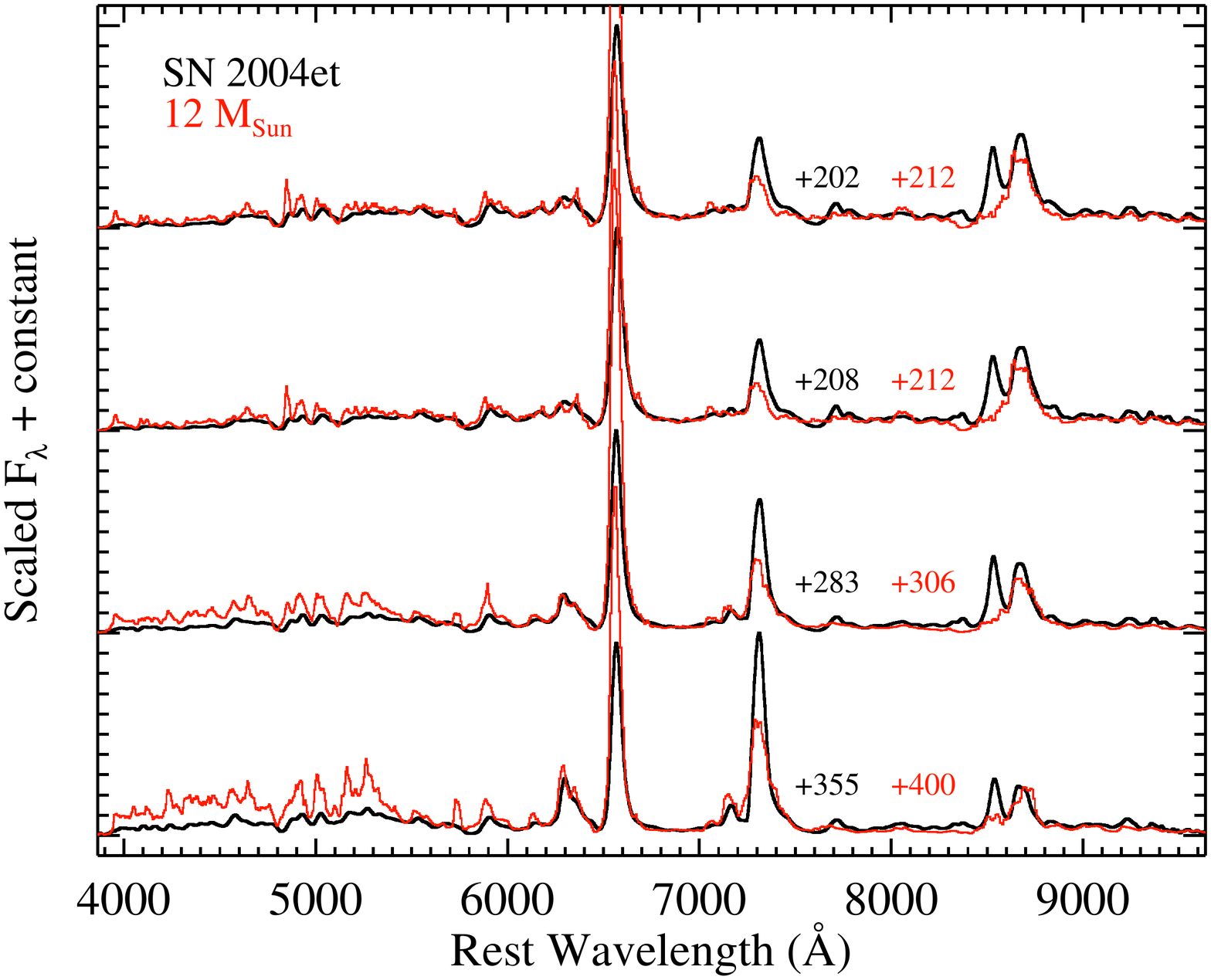} \\
\includegraphics[width=3.6in]{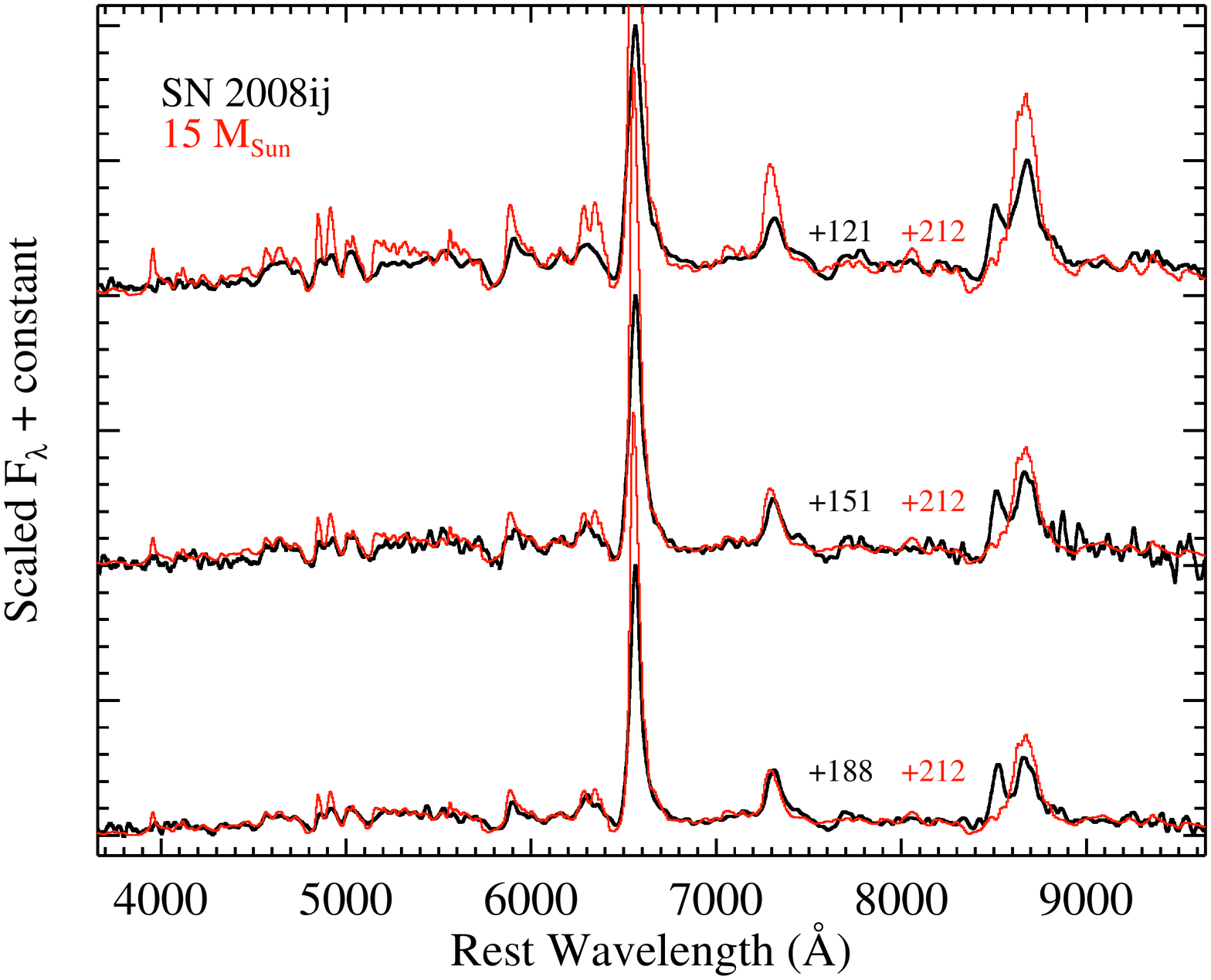} &
\includegraphics[width=3.6in]{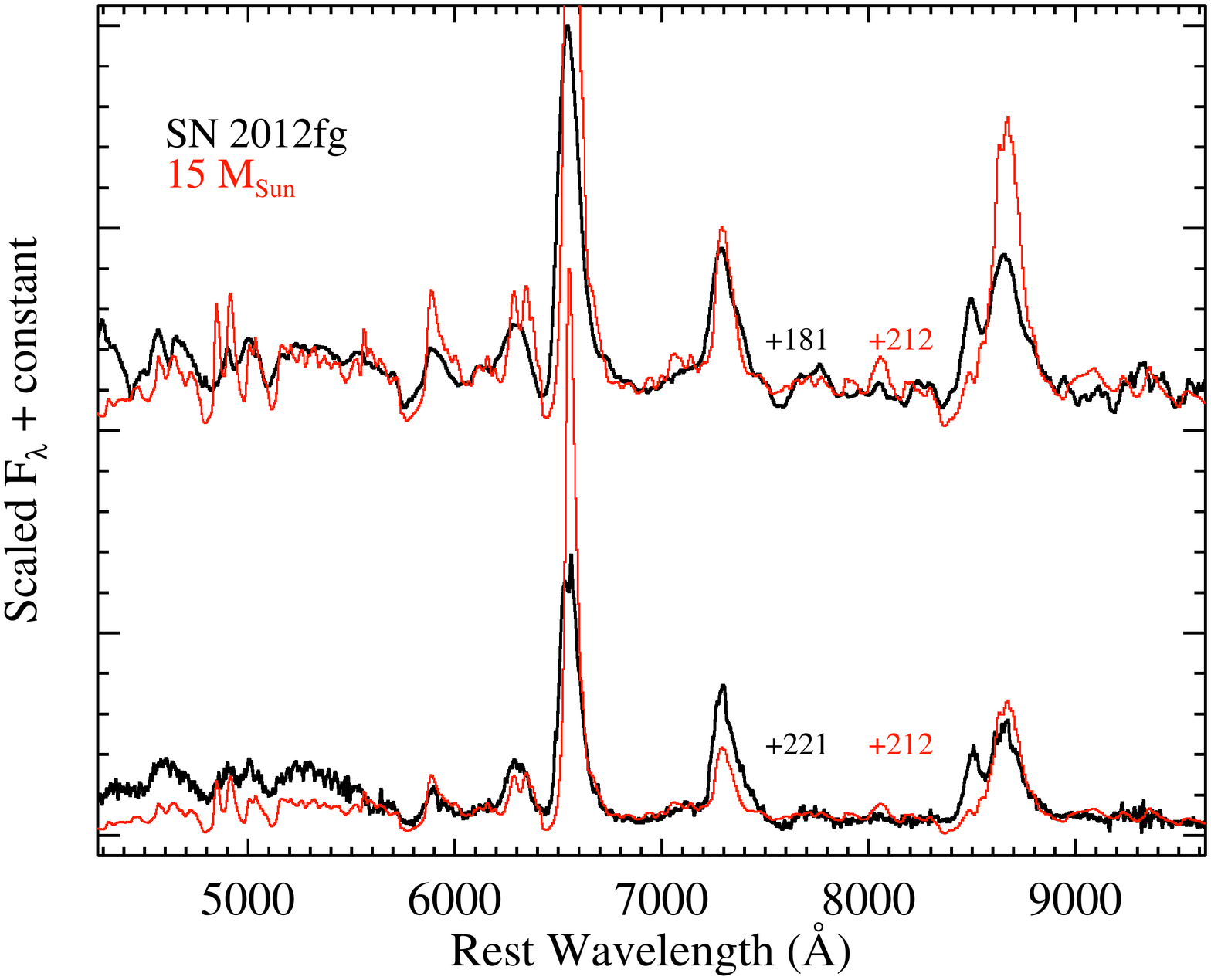} \\
\end{array}$
\caption{A comparison between SNe~IIP and 
  \citet{Jerkstrand14} models. SN~2013am (upper left) is a
  narrow-lined SN~IIP and somewhat resembles the 12~\msun\ model, even
  though its emission features are much narrower than those in the
  model. SN~2012aw (upper right) also matches best to the 12~\msun\
  model and was discussed by 
  \citet{Jerkstrand14}. SNe~2004A and 2004et (middle row) are
  consistent with the 12~\msun\ model, while SNe~2008ij and 2012fg
  (bottom row) are consistent with the 15~\msun\ model. Each spectrum
  is labeled with 
  its rest-frame age relative to explosion (black) and the age of the
  overplotted model spectrum relative to explosion (red). All data have
  been corrected for host-galaxy recession velocity and Galactic
  reddening using the values listed in
  Table~\ref{t:objects}.}\label{f:jerkstrand_models} 
\end{figure*}

SNe~2005cs and 2013am, as mentioned in
Section~\ref{ss:phot}, have some of the lowest HWHM values
(i.e., narrowest emission lines) in our sample and neither are fit very
well by any of the \citet{Jerkstrand14} models (see 
the upper-left panel in Figure~\ref{f:jerkstrand_models}). We do note,
however, that a different suite of models with narrower emission lines
was produced by the same group and presented by \citet{Maguire12}; they
match SN~2005cs, as well as other narrow-lined SNe~IIP, very well.

Using the \citet{Jerkstrand14} models, we find that most of the
objects in our sample (30) resemble their 12~\msun\ model. The  
other eight objects in our sample matched slightly better with their 15~\msun\
model. There were a few individual spectra that were consistent with 
19~\msun\ and 25~\msun\ models, but they usually also resembled 
the 15~\msun\ model. Therefore, the comparisons of our observations to
the models of \citet{Jerkstrand14} appear to support the RSG problem
\citep{Smartt09} in that our SNe~IIP tend to prefer progenitors with
masses smaller than \about16~\msun. A literature search yielded
observed progenitor masses for 13 of the objects in our sample and all
of them are consistent with our findings, having masses in the range
8--18~\msun\
\citep{Smartt09,Vandyk12,Maund13,Tomasella13,Maund14,Bose15}. 
This result has also been found by \citet{Jerkstrand15} using a sample
of 12 SNe~IIP. We
caution, however, that this result is complicated by the fact that
\citet{deMink14} find that \about19~per~cent of apparently single,
massive stars actually come from mergers. Observational signatures of
such systems may be found in future analyses of late-time SN~IIP
spectra, but few predictions currently exist.

As mentioned above, the upper-left panel of
Figure~\ref{f:jerkstrand_models} shows the narrow-lined SN~2013am and
the 12~\msun\ \citet{Jerkstrand14} model. While numerous
emission features are present in both the data and the model, and
the relative strengths of many features match quite well, the emission
lines in the model are significantly broader than those in the
data. The upper-right panel of Figure~\ref{f:jerkstrand_models}
displays SN~2012aw and its best-matching model from
\citet{Jerkstrand14}, again the 12~\msun\ model. The middle row of
Figure~\ref{f:jerkstrand_models} shows two 
more objects, SNe~2004A and 2004et, that are fit well by the 12~\msun\
model, and the bottom row contains spectra of SNe~2008ij and 2012fg,
two of the relatively few objects that are more consistent with their 
15~\msun\ model. 

\begin{figure*}
\centering$
\begin{array}{cc}
\includegraphics[width=3.6in]{sn2012aw_12_model} &
\includegraphics[width=3.6in]{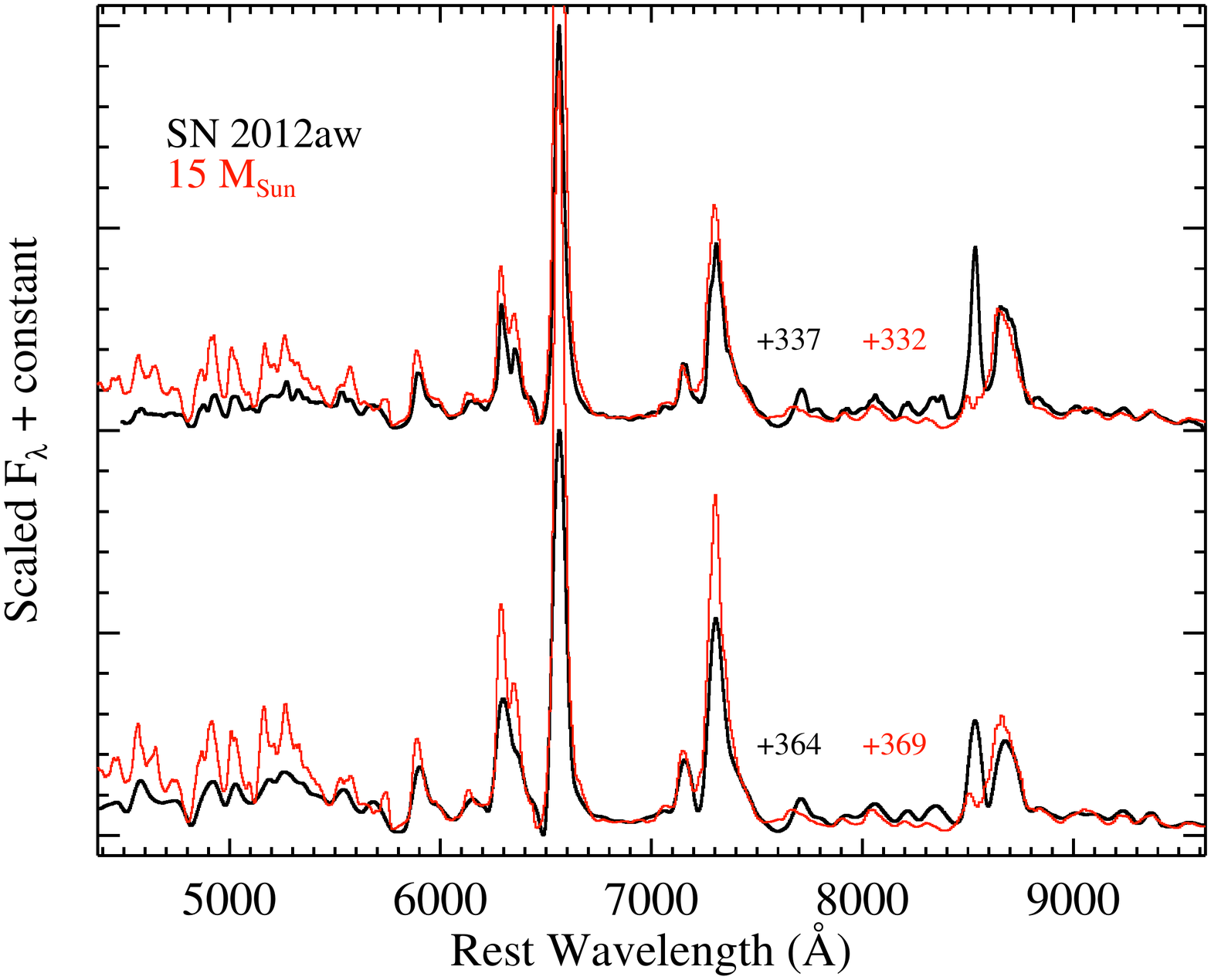} \\
\includegraphics[width=3.6in]{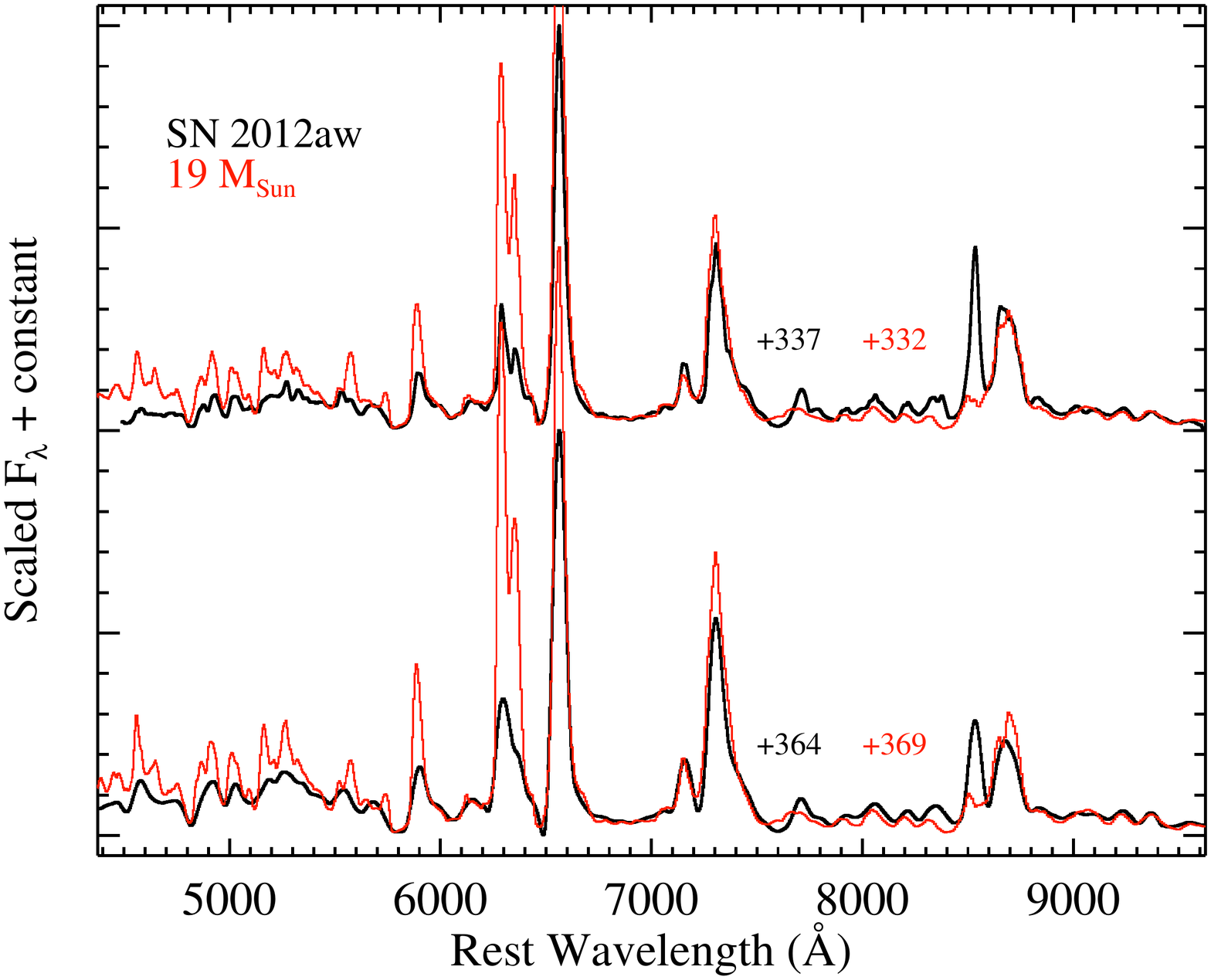} &
\includegraphics[width=3.6in]{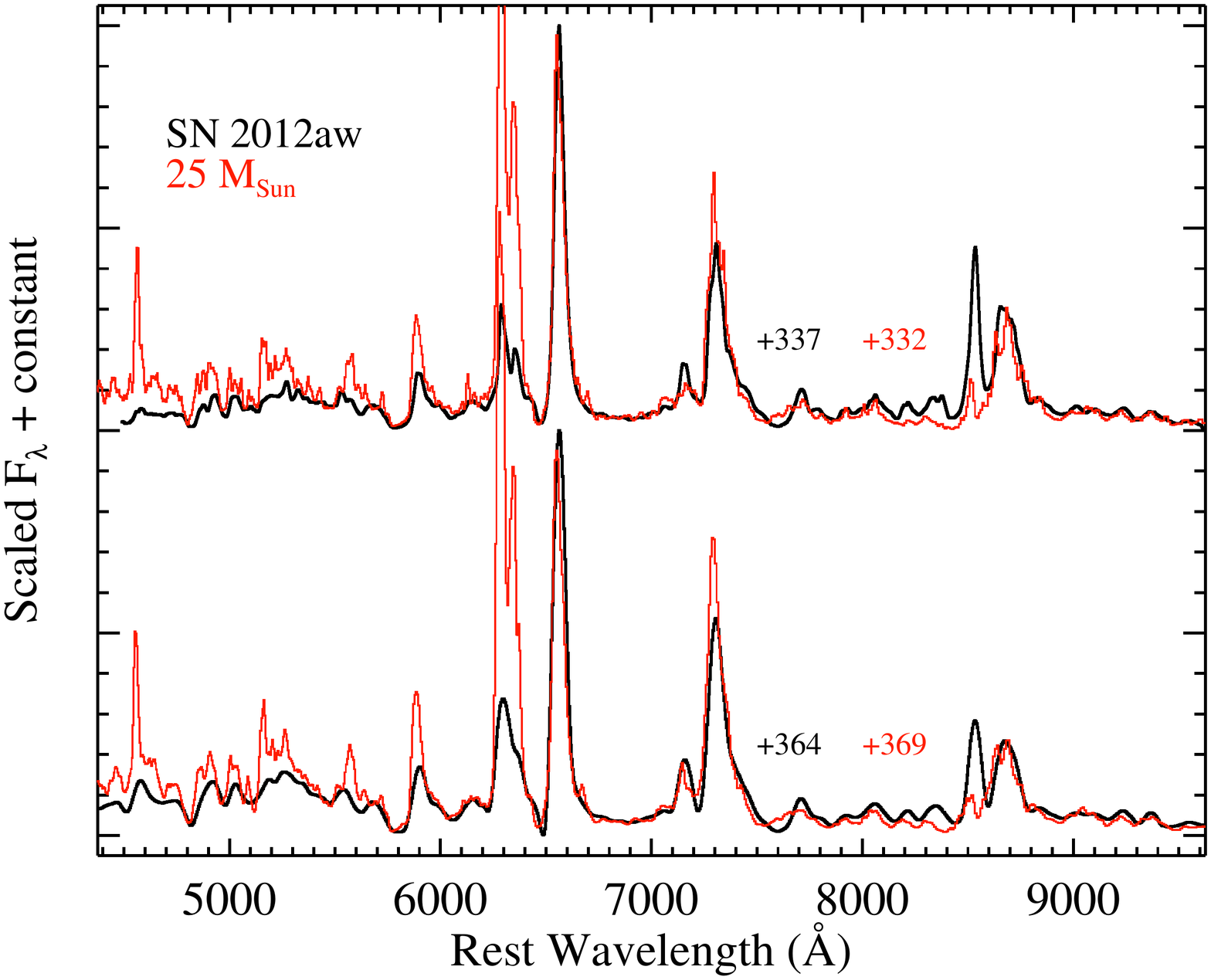} \\
\end{array}$
\caption{A comparison between SN~2012aw and models of four different
  progenitor masses from \citet{Jerkstrand14}. SN~2012aw matches best
  to the 12~\msun\ model (top-left panel) and was discussed by
  \citet{Jerkstrand14}. Other progenitor masses 
  include 15~\msun\ (top-right panel), 19~\msun\ (bottom-left panel),
  and 25~\msun\ (bottom-right panel). Each spectrum is labeled with 
  its rest-frame age relative to explosion (black) and the age of the
  overplotted model spectrum relative to explosion (red). All data have
  been corrected for host-galaxy recession velocity and Galactic
  reddening using the values listed in
  Table~\ref{t:objects}.}\label{f:jerkstrand_models_all} 
\end{figure*}

To highlight the variance caused by different progenitor masses in the
model spectra from \citet{Jerkstrand14}, we present our late-time
spectra of SN~2012aw overplotted with each of their four progenitor
mass models in Figure~\ref{f:jerkstrand_models_all}. The top-left
panel shows the 12~\msun\ model, which is the same as what is shown in
the top-right panel of of Figure~\ref{f:jerkstrand_models}. The model
spectra in each of the other panels are less consistent with the
observed spectra of SN~2012aw, especially in the strength of the
[\ion{O}{I}] $\lambda\lambda$6300, 6364 feature.

\section{Summary \& Conclusions}\label{s:conclusions}

In this work we present 91 late-time, nebular spectra of 38 SNe~IIP,
which is the largest dataset of its kind ever analysed in one
study. We have multiple spectra of most of the objects and many of 
the SNe~IIP have not been studied by other researchers at late
times. Furthermore, most of the spectra presented herein are 
previously unpublished. The observations span 103--1229~d relative to
explosion and are found at distances smaller than 69~Mpc, with a
typical distance of \about21~Mpc. In order to determine whether a
spectrum was truly nebular, and thus should be included in our datset,
we required that the [\ion{O}{I}] 
$\lambda\lambda$6300, 6364 doublet show two distinct peaks that were
$>2\sigma$ above the local continuum. In one case, detection of two
distinct peaks of [\ion{Ca}{II}] $\lambda\lambda$7291, 7324 was allowed
instead. 

We also gathered photometric data for most of the objects in our
sample, including \mplat\ values for every object and \mpk\ values for
nearly two-thirds of them. Also studied were 77 late-time
optical light curves of 25 SNe~IIP, which allowed us to scale spectra of
these objects to interpolated/extrapolated photometry in order to put
them on an accurate absolute flux scale. The late-time linear decline
rates in multiple bands were measured from these data.

In each spectrum we searched for various permitted and forbidden
emission lines from hydrogen, helium, oxygen, magnesium, potassium,
calcium, and iron. 
The resulting measurements can be found in
Tables~\ref{t:hydrogen}--\ref{t:iron}. 

Overall, our spectral feature measurements are consistent with
previous work on individual SNe~IIP and relatively small samples of
objects. The \lpk\ and \ltot\ values of the bluer spectral features
investigated 
in this work tend to be higher for SNe~IIP with brighter \mplat,
possibly indicating a positive correlation between progenitor radius
and mass of $^{56}$Ni produced. The strongest and most robust result
we found is that \ltot\ and \lpk\ values for all spectral features
(except those of helium) tend to be higher for steeper late-time
$V$-band and pseudobolometric slopes (and HWHM values are bigger for
steeper $V$-band 
slopes for the strongest lines as well). A steep late-time $V$-band
slope likely arises from less efficient trapping of $\gamma$-rays and
positrons, which could be caused by multidimensional effects such as
clumping of the ejecta or asphericity of the explosion
itself. Assuming that $\gamma$-rays and positrons can escape
relatively easily, photons should be able to as well via the observed
emission lines, leading to more-luminous spectral features.

The \vpk\ of \hal\ appears mostly blueshifted and approaches zero
velocity at later epochs, and is found to be anticorrelated with the
late-time $V$-band slope. HWHM values for all spectral features
studied tend to decrease with time and median HWHM values are found to
be larger for objects with steeper late-time $V$-band slopes. In addition, the
HWHM of \hal\ and all oxygen spectral features are larger for SNe~IIP
with brighter \mpk\ and \mplat. These observations imply that SNe~IIP
with larger progenitor stars should also have ejecta with a more
physically extended oxygen layer that is well mixed 
with the hydrogen layer \citep[e.g.,][]{Dessart13}.

Various spectral flux ratios are also calculated and investigated
herein. We find the peak flux ratio of the [\ion{O}{I}]
$\lambda\lambda$6300, 6364 doublet to mostly decrease from \about1.0 to
\about0.4–-0.5 as the SN~IIP ejecta transition from optically thick
to optically thin. Also, the ratio of each of the [\ion{O}{I}]
$\lambda\lambda$6300, 6364, [\ion{Ca}{II}] $\lambda\lambda$7291, 7324,
and [\ion{Fe}{II}] $\lambda\lambda$7155, 7172 doublets to \hal\ shows
the same general trend of roughly constant values for $t \la
200$--250~d and then increasing with time thereafter. 

The overall appearance of the shape of each measured spectral profile
was also investigated. The vast majority of emission lines were found
to be single-peaked and none were seen to have flat tops, similar to
what was found by \citet{Maguire12}. We found 8 SNe~IIP showing \hal\
profiles 
with blueshifted peaks and a red shoulder, while three objects had the
opposite asymmetric profile, all at epochs earlier than 300~d past
explosion. One object (SN~2009ls) is included in both of those categories;
it evolves from the former case to the latter between
our two observations. These profile shapes are possibly caused by
asymmetric $^{56}$Ni ejection, likely bipolar in shape
\citep[e.g.,][]{Chugai05}. 

Lastly, comparisons were made to theoretical late-time spectral
models of SNe~IIP from \citet{Dessart13} and \citet{Jerkstrand14}. 
Most of the objects in our sample were consistent with the
\citet{Dessart13} models with relatively low 
metallicity ($Z \le 0.01$). When comparing our dataset to the models
of \citet{Jerkstrand14}, 30 SNe~IIP were most similar to their
12~\msun\ model, while the other 8 objects were better matched by the
15~\msun\ model. This seems to support the RSG problem
\citep{Smartt09} and is consistent with direct observations of the
progenitors of some of the SNe~IIP in our sample.

Although the current sample constitutes the largest late-time SN~IIP
spectral dataset ever studied, it still contains relatively few
objects and only a handful with spectra at more than two
epochs. Future analyses similar to the one undertaken herein would
benefit greatly by expanding the total sample of nebular SN~IIP
spectra. The relatively low luminosity of SNe~IIP at late times makes
obtaining such spectra difficult, even in the era of 10~m
telescopes. Thus, the upcoming 30-m-class telescopes (GMT, E-ELT, and
TMT) will be key to extending this work. Discovering
nearby SNe~IIP in greater numbers and in a wider variety of
host-galaxy types would also be beneficial. 
Large, ``untargeted'' transient searches coming online soon, such as
the Zwicky Transient Factory \citep[ZTF;][]{Bellm14,Smith14} and LSST 
\citep{LSST}, should be able to find most of the nearby
SNe~IIP.

\section*{Acknowledegments}

We would like to thank the referee, in addition
to J.\ Anderson, A.\ Clocchiatti, L.\ Dessart, A.\ Jerkstrand, K.\
Maguire, A.\ Piro, I.\ Shivvers, J.\ Spyromilio, and S.\ Valenti, for
helpful discussions that helped improve this paper. We are also
indebted to many observers and data reducers, 
especially M.\ Childress, B.\ Cobb, O.\ Fox, M.\ Ganeshalingam, L.\ Ho,
I.\ Kleiser, F.\ Serduke, I.\ Shivvers, T.\ Steele, B.\ Tucker, D.\ Wong,
and W.\ Zhang,  
%
as well as the staffs at the Lick, Keck, McDonald, and
Siding Spring Observatories, who made this work possible.
Research at Lick Observatory is partially supported by a generous gift 
from Google.

The HET is a joint project of the University of Texas at Austin, the
Pennsylvania State University, Stanford University,
Ludwig-Maximilians-Universit\"{a}t M\"{u}nchen, and
Georg-August-Universit\"{a}t G\"{o}ttingen. The HET is named in honor
of its principal benefactors, William P.\ Hobby and Robert
E.\ Eberly. The Marcario Low Resolution Spectrograph is named for Mike
Marcario of High Lonesome Optics who fabricated several optics for the
instrument but died before its completion. The LRS is a joint project
of the HET partnership and the Instituto de Astronom\'{i}a de la
Universidad Nacional Aut\'{o}noma de M\'{e}xico. 

Some of the data presented herein were obtained at the W.\ M.\ Keck
Observatory, which is operated as a scientific partnership among the
California Institute of Technology, the University of California, and
NASA; the observatory was made possible by the generous financial
support of the W.\ M.\ Keck Foundation. The authors wish to recognise
and acknowledge the very significant cultural role and reverence that
the summit of Mauna Kea has always had within the indigenous Hawaiian
community; we are most fortunate to have the opportunity to conduct
observations from this mountain.

This research has made use of the NASA/IPAC Extragalactic Database
(NED) which is operated by the Jet Propulsion Laboratory, California
Institute of Technology, under contract with NASA.

J.M.S.\ is supported by an NSF Astronomy and Astrophysics Postdoctoral
Fellowship under award AST-1302771.
J.C.W.'s supernova group at UT Austin is supported by NSF Grant AST
11-09801. 
J.V.\ is supported by Hungarian OTKA Grant NN 107637.
A.V.F.'s group at UC Berkeley has been supported by Gary \& Cynthia               
Bengier, the Richard \& Rhoda Goldman Fund, the Christopher                   
R. Redlich Fund, the TABASGO Foundation, and NSF grant AST-1211916.
R.J.F.\ is supported in part by NSF grant AST-1518052 and from
fellowships from the Alfred P.\ Sloan Foundation and the David and
Lucile Packard Foundation.

\bibliographystyle{mn2e}
\bibliography{/Users/jsilverman/iip_late/astro_refs.bib}


\appendix
\newpage
\onecolumn
\section{Tables of Objects, Spectra, Spectral Measurements, and Photometry}\label{a:tables}

\setlength{\tabcolsep}{4pt}
\scriptsize
\begin{center}
\begin{longtable}{lccccccccc}
\caption{Summary of SNe~IIP\label{t:objects}} \\[-2ex]
\hline \hline
SN Name & Host   & $cz$                       & Distance           & $E\left(B-V\right)_\textrm{MW}$ & MJD of    & Approx. MJD               & $M_\textrm{plat}(V)$     & $M_\textrm{pk}(V)$        & Reference(s)$^\textrm{d}$ \\
        & Galaxy & (km s$^{-1}$)$^\textrm{a}$ & (Mpc)$^\textrm{a}$ & (mag)$^\textrm{b}$     & Discovery & of Explosion$^\textrm{c}$ & (mag)$^\textrm{c}$ & (mag)$^\textrm{c}$ & \\
\hline
\endfirsthead

\multicolumn{10}{c}{{\tablename} \thetable{} --- Continued} \\
\hline \hline
SN Name & Host   & $cz$                       & Distance           & $E\left(B-V\right)_\textrm{MW}$ & MJD of    & Approx. MJD               & $M_\textrm{plat}(V)$      & $M_\textrm{pk}(V)$         & Reference(s)$^\textrm{d}$ \\
        & Galaxy & (km s$^{-1}$)$^\textrm{a}$ & (Mpc)$^\textrm{a}$ & (mag)$^\textrm{b}$     & Discovery & of Explosion$^\textrm{c}$ & (mag)$^\textrm{c}$ & (mag)$^\textrm{c}$ & \\
\hline
\endhead

\hline \hline
\multicolumn{10}{l}{Continued on next page} \\
\endfoot

\hline \hline
\endlastfoot

SN 1988A & NGC 4579 & 1520 & 19.58 & 0.036 & 47177 & 47176 (2)$\phantom{0}$ & $-16.9$ (0.1) & $-17.8$ (0.3) & 1 \\
SN 1988H & NGC 5878 & 1991 & 30.36 & 0.126 & 47224 & 47203 (2)$\phantom{0}$ & $-16.7$ (0.5) & $-16.8$ (0.5) & 2 \\
SN 1989L & NGC 7339 & 1313 & 22.97 & 0.033 & 47679 & 47650 (2)$\phantom{0}$ & $-16.3$ (0.5) &  $\cdots$ & 3,4 \\
SN 1990E & NGC 1035 & 1241 & 17.43 & 0.022 & 47938 & 47935 (2)$\phantom{0}$ & $-16.6$ (0.4) & $-16.7$ (0.4) & 5 \\
SN 1990H & NGC 3294 & 1586 & 29.74 & 0.018 & 47991 & 47918 (2)$\phantom{0}$ & $-15.7$ (0.5) &  $\cdots$ & 6 \\
SN 1990K & NGC 150 & 1583 & 21.01 & 0.012 & 48037 & 48017 (5)$\phantom{0}$ & $-17.2$ (0.5) & $-17.2$ (0.5) & 7 \\
SN 1992ad & NGC 4411 & 1280 & 16.80 & 0.026 & 48805 & 48804 (2)$\phantom{0}$ & $-16.1$ (0.5) &  $\cdots$ & 8 \\
SN 1992bt & NGC 3780 & 2398 & 42.62 & 0.012 & 48976 & 48956 (20) & $-16.6$ (0.5) &  $\cdots$ & 9 \\
\end{longtable}
\end{center}
\normalsize
\twocolumn
\setlength{\tabcolsep}{6pt}

\setlength{\tabcolsep}{4pt}
\onecolumn
\begin{center}
\begin{longtable}{lcccccc}
\caption{Summary of SNe~IIP Spectra\label{t:spec}} \\[-2ex]
\hline \hline
SN Name & MJD$^\textrm{a}$ & Age Since              & Instrument$^\textrm{c}$ & Wavelength  & Resolution$^\textrm{d}$ & Previous    \\
        &                  & Discovery$^\textrm{b}$ &                         & Range (\AA) & (\AA)                   & Publication \\
\hline
\endfirsthead

\multicolumn{7}{c}{{\tablename} \thetable{} --- Continued} \\
\hline \hline
SN Name & MJD$^\textrm{a}$ & Age Since              & Instrument$^\textrm{c}$ & Wavelength  & Resolution$^\textrm{d}$ & Previous    \\
        &                  & Discovery$^\textrm{b}$ &                         & Range (\AA) & (\AA)                   & Publication \\
\hline
\endhead

\hline \hline
\multicolumn{7}{l}{Continued on next page} \\
\endfoot

\hline \hline
\endlastfoot

SN 1988A & 47339.262 & 162 & UVSchmidt & 3482--9253$\phantom{1}$ & 12 & $\cdots$ \\
SN 1988A & 47359.195 & 182 & UVSchmidt & 5940--9104$\phantom{1}$ & 12 & $\cdots$ \\
\hline
SN 1988H & 47343.245 & 119 & UVSchmidt & 6072--9239$\phantom{1}$ & 12 & $\cdots$ \\
\hline
SN 1989L & 47861.000 & 182 & UVSchmidt & 4042--9060$\phantom{1}$ & 12 & $\cdots$ \\
\hline
SN 1990E & 48089.000 & 151 & UVSchmidt & 3936--7023$\phantom{1}$ & 12 & $\cdots$ \\
SN 1990E & 48103.479 & 165 & UVSchmidt & 3859--7001$\phantom{1}$ & 12 & \citet{Schmidt93} \\
SN 1990E & 48133.000 & 195 & UVSchmidt & 6692--9809$\phantom{1}$ & 12 & $\cdots$ \\
SN 1990E & 48242.214 & 303 & UVSchmidt & 5776--8913$\phantom{1}$ & 12 & $\cdots$ \\
\end{longtable}
\end{center}
\twocolumn

\setlength{\tabcolsep}{4pt}
\onecolumn
\begin{center}
\begin{longtable}{lccccc}
\caption{Spectral Feature Measurements of Hydrogen\label{t:hydrogen}} \\[-2ex]
\hline \hline
SN Name & Age Since              & \multirow{2}{*}{$\log\left(\frac{L_\textrm{tot}}{\textrm{erg s}^{-1}}\right)$} & \multirow{2}{*}{$\log\left(\frac{L_\textrm{pk}}{\textrm{erg s}^{-1}\textrm{ \AA}^{-1}}\right)$}               & \vpk              & HWHM \\
        & Discovery$^\textrm{a}$ &                                                                                &                                                                                                               & (100 km s$^{-1}$) & (1000 km s$^{-1}$) \\
\hline
\endfirsthead

\multicolumn{6}{c}{{\tablename} \thetable{} --- Continued} \\
\hline \hline
SN Name & Age Since              & \multirow{2}{*}{$\log\left(\frac{L_\textrm{tot}}{\textrm{erg s}^{-1}}\right)$} & \multirow{2}{*}{$\log\left(\frac{L_\textrm{pk}}{\textrm{erg s}^{-1}\textrm{ \AA}^{-1}}\right)$}               & \vpk              & HWHM \\
        & Discovery$^\textrm{a}$ &                                                                                &                                                                                                               & (100 km s$^{-1}$) & (1000 km s$^{-1}$) \\
\hline
\endhead

\hline \hline
\multicolumn{6}{l}{Continued on next page} \\
\endfoot

\hline \hline
\endlastfoot

\multicolumn{6}{c}{\hal} \\
\hline

SN 1988A & $\phantom{1}$162 & 40.19 (0.03) & 38.38 (0.02) & $\phantom{0}$$\phantom{-}$0.40 (0.05) & 1.45 (0.04) \\ 
SN 1988A & $\phantom{1}$182 & 39.77 (0.04) & 37.98 (0.03) & $\phantom{0}$$\phantom{-}$0.43 (0.05) & 1.35 (0.03) \\ 
SN 1988H & $\phantom{1}$119 & 40.05 (0.02) & 38.06 (0.02) & $\phantom{0}$$-$2.52 (0.05) & 2.35 (0.04) \\ 
SN 1989L & $\phantom{1}$182 & 39.59 (0.06) & 37.82 (0.04) & $\phantom{0}$$\phantom{-}$3.39 (0.17) & 1.36 (0.04) \\ 
SN 1990E & $\phantom{1}$151 & 39.49 (0.09) & 37.37 (0.06) & $\phantom{0}$$-$0.09 (0.05) & 3.30 (0.06) \\ 
SN 1990E & $\phantom{1}$165 & 39.31 (0.11) & 37.18 (0.06) & $\phantom{0}$$-$1.22 (0.05) & 3.32 (0.07) \\ 
SN 1990E & $\phantom{1}$304 & 39.07 (0.06) & 37.21 (0.04) & $\phantom{0}$$-$0.53 (0.05) & 1.67 (0.03) \\
\end{longtable}
\end{center}
\twocolumn

\setlength{\tabcolsep}{4pt}
\onecolumn
\begin{center}
\begin{longtable}{lccccc}
\caption{Spectral Feature Measurements of Helium\label{t:helium}} \\[-2ex]
\hline \hline
SN Name & Age Since              & \multirow{2}{*}{$\log\left(\frac{L_\textrm{tot}}{\textrm{erg s}^{-1}}\right)$} & \multirow{2}{*}{$\log\left(\frac{L_\textrm{pk}}{\textrm{erg s}^{-1}\textrm{ \AA}^{-1}}\right)$}               & \vpk              & HWHM \\
        & Discovery$^\textrm{a}$ &                                                                                &                                                                                                               & (100 km s$^{-1}$) & (1000 km s$^{-1}$) \\
\hline
\endfirsthead

\multicolumn{6}{c}{{\tablename} \thetable{} --- Continued} \\
\hline \hline
SN Name & Age Since              & \multirow{2}{*}{$\log\left(\frac{L_\textrm{tot}}{\textrm{erg s}^{-1}}\right)$} & \multirow{2}{*}{$\log\left(\frac{L_\textrm{pk}}{\textrm{erg s}^{-1}\textrm{ \AA}^{-1}}\right)$}               & \vpk              & HWHM \\
        & Discovery$^\textrm{a}$ &                                                                                &                                                                                                               & (100 km s$^{-1}$) & (1000 km s$^{-1}$) \\
\hline
\endhead

\hline \hline
\multicolumn{6}{l}{Continued on next page} \\
\endfoot

\hline \hline
\endlastfoot

\multicolumn{6}{c}{\ion{He}{I} $\lambda$5876} \\
\hline

SN 1988A & $\phantom{1}$162 & 39.49 (0.04) & 37.81 (0.03) & $\phantom{0}$$\phantom{-}$3.54 (0.18) & 1.43 (0.03) \\ 
SN 1989L & $\phantom{1}$182 & 38.95 (0.05) & 37.05 (0.03) & $\phantom{0}$$\phantom{-}$9.70 (0.48) & 2.91 (0.03) \\ 
SN 1990E & $\phantom{1}$151 & 38.80 (0.05) & 36.79 (0.04) & $\phantom{-}$10.73 (0.54) & 4.38 (0.08) \\ 
SN 1990E & $\phantom{1}$165 & 38.71 (0.03) & 36.61 (0.02) & $\phantom{0}$$\phantom{-}$2.79 (0.14) & 5.42 (0.14) \\ 
SN 1990E & $\phantom{1}$304 & 37.99 (0.16) & 36.27 (0.14) & $\phantom{0}$$\phantom{-}$3.51 (0.18) & 1.99 (0.09) \\ 
SN 1990H & $\phantom{1}$193 & 39.08 (0.10) & 37.52 (0.09) & $-$17.23 (0.05) & 1.34 (0.03) \\ 
SN 1990K & $\phantom{1}$205 & 38.51 (0.21) & 36.95 (0.15) & $-$19.37 (0.05) & 0.92 (0.02) \\
\end{longtable}
\end{center}
\twocolumn

\setlength{\tabcolsep}{4pt}
\onecolumn
\begin{center}
\begin{longtable}{lcccccc}
\caption{Spectral Feature Measurements of Oxygen\label{t:oxygen}} \\[-2ex]
\hline \hline
SN Name & Age Since              & \multirow{2}{*}{$\log\left(\frac{L_\textrm{tot}}{\textrm{erg s}^{-1}}\right)$} & \multirow{2}{*}{$\log\left(\frac{L_\textrm{pk}}{\textrm{erg s}^{-1}\textrm{ \AA}^{-1}}\right)$}               & \vpk$^\textrm{b}$ & HWHM$_1$$^\textrm{c}$ & HWHM$_2$$^\textrm{d}$ \\
        & Discovery$^\textrm{a}$ &                                                                                &                                                                                                               & (100 km s$^{-1}$) & (1000 km s$^{-1}$)    & (1000 km s$^{-1}$)\\
\hline
\endfirsthead

\multicolumn{7}{c}{{\tablename} \thetable{} --- Continued} \\
\hline \hline
SN Name & Age Since              & \multirow{2}{*}{$\log\left(\frac{L_\textrm{tot}}{\textrm{erg s}^{-1}}\right)$} & \multirow{2}{*}{$\log\left(\frac{L_\textrm{pk}}{\textrm{erg s}^{-1}\textrm{ \AA}^{-1}}\right)$}               & \vpk$^\textrm{b}$ & HWHM$_1$$^\textrm{c}$ & HWHM$_2$$^\textrm{d}$ \\
        & Discovery$^\textrm{a}$ &                                                                                &                                                                                                               & (100 km s$^{-1}$) & (1000 km s$^{-1}$)    & (1000 km s$^{-1}$)\\
\hline
\endhead

\hline \hline
\multicolumn{7}{l}{Continued on next page} \\
\endfoot

\hline \hline
\endlastfoot

\multicolumn{7}{c}{[\ion{O}{I}] $\lambda\lambda$6300,6364} \\
\hline

SN 1988A & $\phantom{1}$162 & 39.65 (0.03) & 37.64 (0.03) & $\phantom{0}$$-$0.17 (0.05) & 2.85 (0.18) & 1.48 (0.08) \\ 
SN 1988A & $\phantom{1}$182 & 39.24 (0.04) & 37.20 (0.05) & $\phantom{0}$$-$0.20 (0.05) & 2.62 (0.25) & 1.53 (0.10) \\ 
SN 1988H & $\phantom{1}$119 & 39.34 (0.04) & 37.37 (0.04) & $\phantom{0}$$-$5.80 (0.05) & 1.82 (0.07) & 1.28 (0.05) \\ 
SN 1989L & $\phantom{1}$182 & 39.06 (0.05) & 37.09 (0.06) & $\phantom{0}$$-$2.86 (0.05) & 2.03 (0.05) & 1.35 (0.06) \\ 
SN 1990E & $\phantom{1}$151 & 38.80 (0.09) & 36.75 (0.08) & $\phantom{0}$$-$5.12 (0.05) & 2.42 (0.14) & 1.73 (0.09) \\ 
SN 1990E & $\phantom{1}$165 & 38.75 (0.04) & 36.63 (0.03) & $\phantom{0}$$\phantom{-}$1.15 (0.06) & 4.12 (0.67) & 1.72 (0.46) \\ 
SN 1990E & $\phantom{1}$304 & 38.54 (0.06) & 36.66 (0.05) & $\phantom{0}$$-$2.22 (0.05) & 1.39 (0.04) & 0.91 (0.05) \\
\end{longtable}
\end{center}
\twocolumn

\setlength{\tabcolsep}{4pt}
\onecolumn
\begin{center}
\begin{longtable}{lccccc}
\caption{Spectral Feature Measurements of Magnesium\label{t:magnesium}} \\[-2ex]
\hline \hline
SN Name & Age Since              & \multirow{2}{*}{$\log\left(\frac{L_\textrm{tot}}{\textrm{erg s}^{-1}}\right)$} & \multirow{2}{*}{$\log\left(\frac{L_\textrm{pk}}{\textrm{erg s}^{-1}\textrm{ \AA}^{-1}}\right)$}               & \vpk              & HWHM \\
        & Discovery$^\textrm{a}$ &                                                                                &                                                                                                               & (100 km s$^{-1}$) & (1000 km s$^{-1}$) \\
\hline
\endfirsthead

\multicolumn{6}{c}{{\tablename} \thetable{} --- Continued} \\
\hline \hline
SN Name & Age Since              & \multirow{2}{*}{$\log\left(\frac{L_\textrm{tot}}{\textrm{erg s}^{-1}}\right)$} & \multirow{2}{*}{$\log\left(\frac{L_\textrm{pk}}{\textrm{erg s}^{-1}\textrm{ \AA}^{-1}}\right)$}               & \vpk              & HWHM \\
        & Discovery$^\textrm{a}$ &                                                                                &                                                                                                               & (100 km s$^{-1}$) & (1000 km s$^{-1}$) \\
\hline
\endhead

\hline \hline
\multicolumn{6}{l}{Continued on next page} \\
\endfoot

\hline \hline
\endlastfoot

\multicolumn{6}{c}{\ion{Mg}{I}] $\lambda$4571} \\
\hline

SN 1992ad & $\phantom{1}$225 & 38.36 (0.04) & 36.78 (0.04) & $\phantom{0}$$\phantom{-}$9.35 (0.47) & 3.88 (0.15) \\ 
SN 1992ad & $\phantom{1}$286 & 38.15 (0.05) & 36.47 (0.05) & $\phantom{0}$$-$5.13 (0.07) & 3.78 (0.11) \\ 
SN 1992H & $\phantom{1}$203 & 39.11 (0.08) & 37.37 (0.07) & $\phantom{0}$$-$1.05 (0.07) & 2.76 (0.07) \\ 
SN 1992H & $\phantom{1}$221 & 39.01 (0.03) & 37.33 (0.03) & $\phantom{0}$$\phantom{-}$4.05 (0.20) & 3.06 (0.05) \\ 
SN 1992H & $\phantom{1}$382 & 38.60 (0.06) & 36.93 (0.05) & $\phantom{0}$$-$3.50 (0.07) & 2.07 (0.04) \\ 
SN 1992H & $\phantom{1}$425 & 38.66 (0.06) & 36.66 (0.05) & $\phantom{0}$$-$0.48 (0.07) & 4.77 (0.08) \\ 
SN 1992H & $\phantom{1}$461 & 38.33 (0.11) & 36.50 (0.11) & $\phantom{0}$$\phantom{-}$3.19 (0.16) & 4.64 (0.18) \\
\end{longtable}
\end{center}
\twocolumn

\setlength{\tabcolsep}{4pt}
\onecolumn
\begin{center}
\begin{longtable}{lcccccc}
\caption{Spectral Feature Measurements of Calcium\label{t:calcium}} \\[-2ex]
\hline \hline
SN Name & Age Since              & \multirow{2}{*}{$\log\left(\frac{L_\textrm{tot}}{\textrm{erg s}^{-1}}\right)$} & \multirow{2}{*}{$\log\left(\frac{L_\textrm{pk}}{\textrm{erg s}^{-1}\textrm{ \AA}^{-1}}\right)$}               & \vpk$^\textrm{b}$ & HWHM$_1$$^\textrm{c}$ & HWHM$_2$$^\textrm{d}$ \\
        & Discovery$^\textrm{a}$ &                                                                                &                                                                                                               & (100 km s$^{-1}$) & (1000 km s$^{-1}$)    & (1000 km s$^{-1}$)\\
\hline
\endfirsthead

\multicolumn{7}{c}{{\tablename} \thetable{} --- Continued} \\
\hline \hline
SN Name & Age Since              & \multirow{2}{*}{$\log\left(\frac{L_\textrm{tot}}{\textrm{erg s}^{-1}}\right)$} & \multirow{2}{*}{$\log\left(\frac{L_\textrm{pk}}{\textrm{erg s}^{-1}\textrm{ \AA}^{-1}}\right)$}               & \vpk$^\textrm{b}$ & HWHM$_1$$^\textrm{c}$ & HWHM$_2$$^\textrm{d}$ \\
        & Discovery$^\textrm{a}$ &                                                                                &                                                                                                               & (100 km s$^{-1}$) & (1000 km s$^{-1}$)    & (1000 km s$^{-1}$)\\
\hline
\endhead

\hline \hline
\multicolumn{7}{l}{Continued on next page} \\
\endfoot

\hline \hline
\endlastfoot

\multicolumn{7}{c}{[\ion{Ca}{II}] $\lambda\lambda$7291,7324} \\
\hline

SN 1988A & $\phantom{1}$162 & 39.67 (0.05) & 37.90 (0.04) & $\phantom{0}$$-$2.67 (0.54) & 0.66 (0.02) & 1.12 (0.03) \\ 
SN 1988A & $\phantom{1}$182 & 39.20 (0.05) & 37.45 (0.05) & $\phantom{0}$$-$2.68 (0.54) & 0.72 (0.02) & 0.94 (0.02) \\ 
SN 1988H & $\phantom{1}$119 & 39.34 (0.03) & 37.60 (0.04) & $\phantom{0}$$-$0.82 (0.04) & 0.86 (0.05) & 1.06 (0.08) \\ 
SN 1989L & $\phantom{1}$182 & 39.01 (0.04) & 37.27 (0.04) & $\phantom{0}$$\phantom{-}$0.95 (0.72) & 0.73 (0.03) & 0.97 (0.05) \\ 
SN 1990E & $\phantom{1}$195 & 38.77 (0.05) & 37.01 (0.05) & $\phantom{0}$$\phantom{-}$1.74 (0.09) & 1.65 (0.40) & 1.01 (0.47) \\ 
SN 1990E & $\phantom{1}$304 & 38.57 (0.04) & 36.83 (0.03) & $\phantom{0}$$\phantom{-}$3.47 (0.17) & 1.20 (0.42) & 1.14 (0.59) \\ 
SN 1990H & $\phantom{1}$193 & 39.17 (0.08) & 37.40 (0.09) & $\phantom{0}$$-$7.42 (0.04) & 1.48 (0.25) & 0.79 (0.22) \\
\end{longtable}
\end{center}
\twocolumn

\setlength{\tabcolsep}{4pt}
\onecolumn
\begin{center}
\begin{longtable}{lccccc}
\caption{Spectral Feature Measurements of Iron\label{t:iron}} \\[-2ex]
\hline \hline
SN Name & Age Since              & \multirow{2}{*}{$\log\left(\frac{L_\textrm{tot}}{\textrm{erg s}^{-1}}\right)$} & \multirow{2}{*}{$\log\left(\frac{L_\textrm{pk}}{\textrm{erg s}^{-1}\textrm{ \AA}^{-1}}\right)$}               & \vpk              & HWHM \\
        & Discovery$^\textrm{a}$ &                                                                                &                                                                                                               & (100 km s$^{-1}$) & (1000 km s$^{-1}$) \\
\hline
\endfirsthead

\multicolumn{6}{c}{{\tablename} \thetable{} --- Continued} \\
\hline \hline
SN Name & Age Since              & \multirow{2}{*}{$\log\left(\frac{L_\textrm{tot}}{\textrm{erg s}^{-1}}\right)$} & \multirow{2}{*}{$\log\left(\frac{L_\textrm{pk}}{\textrm{erg s}^{-1}\textrm{ \AA}^{-1}}\right)$}               & \vpk              & HWHM \\
        & Discovery$^\textrm{a}$ &                                                                                &                                                                                                               & (100 km s$^{-1}$) & (1000 km s$^{-1}$) \\
\hline
\endhead

\hline \hline
\multicolumn{6}{l}{Continued on next page} \\
\endfoot

\hline \hline
\endlastfoot

\multicolumn{6}{c}{\ion{Fe}{II} $\lambda$5018} \\
\hline

SN 1988A & $\phantom{1}$162 & 39.30 (0.04) & 37.50 (0.03) & $\phantom{0}$$\phantom{-}$9.13 (0.46) & 2.28 (0.03) \\ 
SN 1992ad & $\phantom{1}$225 & 38.51 (0.04) & 36.83 (0.03) & $\phantom{0}$$-$5.04 (0.06) & 2.18 (0.04) \\ 
SN 1992ad & $\phantom{1}$286 & 38.17 (0.08) & 36.53 (0.06) & $\phantom{0}$$-$3.92 (0.06) & 1.91 (0.04) \\ 
SN 1992H & $\phantom{1}$203 & 39.24 (0.07) & 37.56 (0.05) & $\phantom{0}$$-$3.12 (0.06) & 2.50 (0.04) \\ 
SN 1992H & $\phantom{1}$221 & 39.38 (0.03) & 37.50 (0.02) & $\phantom{0}$$\phantom{-}$0.52 (0.06) & 2.95 (0.03) \\ 
SN 1992H & $\phantom{1}$382 & 38.51 (0.07) & 36.81 (0.06) & $\phantom{0}$$\phantom{-}$3.82 (0.19) & 2.49 (0.09) \\ 
SN 1992H & $\phantom{1}$425 & 38.41 (0.09) & 36.54 (0.09) & $\phantom{0}$$\phantom{-}$2.58 (0.13) & 3.33 (0.28) \\
\end{longtable}
\end{center}
\twocolumn

\onecolumn
\begin{center}
\begin{longtable}{lccccl}
\caption{Late-Time Light Curves\label{t:lc}} \\[-2ex]
\hline \hline
SN Name & Filter & \# Points & MJD range & Slope$^\textrm{a}$ & Photometry Reference \\
\hline
\endfirsthead

\multicolumn{6}{c}{{\tablename} \thetable{} --- Continued} \\
\hline \hline
SN Name & Filter & \# Points & MJD range & Slope$^\textrm{a}$ & Photometry Reference \\
\hline
\endhead

\hline \hline
\multicolumn{6}{l}{Continued on next page} \\
\endfoot

\hline \hline
\endlastfoot

SN 1988A & B & 3 & 47531--47623 & 0.985 (0.124) & \citet{Benetti91} \\
SN 1988A & V & 4 & 47339--47623 & 1.075 (0.023) & \citet{Benetti91} \\
SN 1988A & R & 5 & 47362--47689 & 0.855 (0.036) & \citet{Ruiz-Lapuente90a} \\
\hline
SN 1990E & B & 3 & 48076--48188 & 0.619 (0.326) & \citet{Schmidt93} \\
SN 1990E & V & 10 & 48084--48244 & 0.912 (0.094) & \citet{Schmidt93} \\
SN 1990E & R & 13 & 48084--48272 & 0.889 (0.059) & \citet{Schmidt93} \\
SN 1990E & I & 13 & 48084--48272 & 0.959 (0.054) & \citet{Schmidt93} \\
\hline
SN 1992H & B & 11 & 48808--49105 & 0.757 (0.011) & \citet{Clocchiatti96} \\
\end{longtable}
\end{center}
\twocolumn

\setlength{\tabcolsep}{4pt}
\onecolumn
\begin{center}
\begin{longtable}{lccccccc}
\caption{Interpolated/Extrapolated Photometry at Spectral Epochs\label{t:spec_phot}} \\[-2ex]
\hline \hline
SN Name & MJD$^\textrm{a}$ & Age Since              & $B$   & $V$   & $R$   & $I$   & Unf.$^\textrm{d}$ \\
        &                  & Discovery$^\textrm{b}$ & (mag)$^\textrm{c}$ & (mag)$^\textrm{c}$ & (mag)$^\textrm{c}$ & (mag)$^\textrm{c}$ & (mag)$^\textrm{c}$ \\
\hline
\endfirsthead

\multicolumn{8}{c}{{\tablename} \thetable{} --- Continued} \\
\hline \hline
SN Name & MJD$^\textrm{a}$ & Age Since              & $B$   & $V$   & $R$   & $I$   & Unf.$^\textrm{d}$ \\
        &                  & Discovery$^\textrm{b}$ & (mag)$^\textrm{c}$ & (mag)$^\textrm{c}$ & (mag)$^\textrm{c}$ & (mag)$^\textrm{c}$ & (mag)$^\textrm{c}$ \\
\hline
\endhead

\hline \hline
\multicolumn{8}{l}{Continued on next page} \\
\endfoot

\hline \hline
\endlastfoot

SN 1988A & 47339.262 & 162 & 18.17 (0.09)  & 17.24 (0.05)  & 16.10 (0.09)  & $\cdots$  & $\cdots$  \\
SN 1988A & 47359.195 & 182 & 18.37 (0.09)  & 17.45 (0.05)  & 16.27 (0.09)  & $\cdots$  & $\cdots$  \\
SN 1988H & 47343.245 & 119 & 18.69 (0.09)  & 17.48 (0.04)  & $\cdots$  & $\cdots$  & $\cdots$  \\
SN 1990E & 48089.000 & 151 & 20.48 (0.02)  & 18.92 (0.09)  & 17.55 (0.12)  & 16.70 (0.06)  & $\cdots$  \\
SN 1990E & 48103.479 & 165 & 20.57 (0.02)  & 19.05 (0.09)  & 17.67 (0.12)  & 16.84 (0.06)  & $\cdots$  \\
SN 1990E & 48133.000 & 195 & 20.75 (0.02)  & 19.32 (0.09)  & 17.94 (0.12)  & 17.12 (0.06)  & $\cdots$  \\
SN 1990E & 48242.214 & 303 & 21.43 (0.02)  & 20.32 (0.09)  & 18.91 (0.12)  & 18.17 (0.06)  & $\cdots$  \\
SN 1990K & 48184.356 & 147 & 20.14 (0.10)  & 19.06 (0.08)  & 17.93 (0.11)  & $\cdots$  & $\cdots$  \\
\end{longtable}
\end{center}
\label{lastpage}

\end{document}